\documentclass[11pt,preprint]{aastex}

\received{}
\revised{}
\accepted{}
\journalid{}{}
\articleid{}{}
\paperid{}
\cpright{}{}
\ccc{}

\newcommand{\beq}{\begin{equation}}
\newcommand{\eeq}{\end{equation}}
\newcommand{\tny}[1]{\mbox{\tiny #1}}

\newcommand{\cm}{\mbox{ cm}}
\newcommand{\sr}{\mbox{ sr}}
\newcommand{\se}{\mbox{ s}}
\newcommand{\erg}{\mbox{ erg}}
\newcommand{\Hz}{\mbox{ Hz}}

\newcommand{\MHz}{\mbox{ MHz}}
\newcommand{\GHz}{\mbox{ GHz}}
\newcommand{\pc}{\mbox{ pc}}
\newcommand{\kpc}{\mbox{ kpc}}

\newcommand{\eV}{\mbox{ eV}}

\newcommand{\MeV}{\mbox{ MeV}}
\newcommand{\GeV}{\mbox{ GeV}}
\newcommand{\TeV}{\mbox{ TeV}}
\newcommand{\K}{\mbox{ K}}
\newcommand{\de}{^{\circ}}
\newcommand{\ph}{\mbox{ ph}}
\newcommand{\muG}{\mbox{ $\mu$G}}
\newcommand{\fin}{\mbox{ .}}
\newcommand{\coma}{\mbox{ ,}}
\newcommand{\AND}{\quad\mbox{ and }\quad}

\newcommand{\gama}{$\gamma$}
\newcommand{\HI}{$\mbox{H \small I}$ }
\newcommand{\HII}{$\mbox{H \small II}$ }
\newcommand{\HH}{$\mbox{H}_2$ }
\newcommand{\HImin}{$\mbox{H \small I}$}
\newcommand{\HIImin}{$\mbox{H \small II}$}
\newcommand{\HHmin}{$\mbox{H}_2$}
\newcommand{\Halpha}{$\mbox{H}\alpha$ }
\newcommand{\CO}{$^{12}\mbox{C}^{16}\mbox{O}$ }
\newcommand{\COmin}{$^{12}\mbox{C}^{16}\mbox{O}$}
\newcommand{\Iunits}{\ph\se^{-1}\cm^{-2}\sr^{-1}}

\newcommand{\junits}{\erg\se^{-1}\cm^{-3}\sr^{-1}\Hz^{-1}}
\newcommand{\angst}{\mbox{\AA}}
\newcommand{\mla}{\la}            
\newcommand{\mga}{\ga}            
\newcommand{\CL}{\%\mbox{ C.L.}}  
\newcommand{\at}{\mbox{at a }}    
\newcommand{\atan}{\mbox{at an }} 

\begin{document}

\title{The Case for a Low Extragalactic Gamma-ray Background}

\author{Uri Keshet\altaffilmark{1}, Eli Waxman\altaffilmark{1,4} 
and Abraham Loeb\altaffilmark{2,3}}
\altaffiltext{1}{Physics Faculty, Weizmann Institute, Rehovot 76100, Israel; keshet@wicc.weizmann.ac.il, waxman@wicc.weizmann.ac.il}
\altaffiltext{2}{Astronomy Department, Harvard University, 60 Garden Street, 
Cambridge, MA 02138, USA}
\altaffiltext{3}{Minerva Einstein Center, Weizmann Institute}
\altaffiltext{4}{Incumbent of the Beracha foundation career development chair}



\begin{abstract}

Measurements of the diffuse extragalactic \gama-ray background (EGRB) are 
complicated by a strong Galactic foreground. 
Estimates of the EGRB flux and spectrum, obtained by modeling the Galactic 
emission, have produced a variety of (sometimes conflicting) results. 
The latest analysis of the EGRET data found an isotropic flux 
$I_x=1.45\pm0.05$ above $100\MeV$, in units of $10^{-5} \Iunits$.
We analyze the EGRET data in search for robust constraints on the EGRB flux, 
finding the \gama-ray sky strongly dominated by Galactic foreground even at 
high latitudes, with no conclusive evidence for an additional isotropic 
component. 
The \gama-ray intensity measured towards the Galactic poles is similar to or 
lower than previous estimates of $I_x$, even before Galactic foreground 
subtraction. 
The high latitude profile of the \gama-ray data is disk-like for 
$40\de\la|b|\la70\de$, and even steeper for $|b|\ga70\de$; overall it 
exhibits strong Galactic features and is well fit by a simple Galactic model. 
Based on the $|b|>40\de$ data we find that $I_x < 0.5$ at a $99\%$ confidence 
level, with evidence for a much lower flux. 
We show that correlations with Galactic tracers, previously used to identify 
the Galactic foreground and estimate $I_x$, are not satisfactory; 
the results depend on the tracers used and on the part of the sky examined, 
because the Galactic emission is not linear in the Galactic tracers, and 
exhibits spectral variations across the sky. 
The low EGRB flux favored by our analysis places stringent limits on 
extragalactic scenarios involving \gama-ray emission, such as radiation from 
blazars, intergalactic shocks and production of ultra-high energy cosmic rays 
and neutrinos. 
We suggest methods by which future \gama-ray missions such as GLAST and AGILE 
could indirectly identify the EGRB. 

\end{abstract}



\section{Introduction}
\label{sec:intro}
\indent

Extragalactic \gama-ray astronomy is a rapidly evolving field of research, 
bearing important implications for cosmology and for high energy astrophysics 
(e.g. Sreekumar et al. 1998; Waxman \& Loeb 2000). 
Maps of the high energy ($\ga 100\MeV$) \gama-ray sky have been produced by 
several experiments since the 1960s, with gradually higher resolution and 
lower background noise. 
The main features of such maps are the bright emission from the Galactic plane 
and the strong correlations with maps of various components of the Galaxy, 
indicating that the \gama-ray sky is dominated by emission originating 
from within the Milky Way galaxy. 
Many efforts have been made to estimate the extragalactic contribution to the 
\gama-ray background, which is assumed to be isotropic on large scales.  
However, a separation of the diffuse extragalactic \gama-ray background (EGRB) 
from the strong Galactic foreground has proven to be a difficult task, due to 
the uncertainty of present models for the Galactic \gama-ray emission. 

The intensity and anisotropy of the \gama-ray sky suggests that the 
Galactic \gama-ray emission is mostly produced by interactions of relativistic 
electrons and protons (cosmic rays) with the matter and radiation fields 
within the Milky Way \cite{Fichtel78}.
Thus, the Galactic foreground is modeled based on the Galactic distributions 
of cosmic rays (CRs), of gas (mostly \HImin, \HII and \HHmin), and of 
the radiation fields (mainly optical and infrared). 
These distributions are, in turn, inferred from observations of their 
radiative signature (Galactic tracers), 
such as synchrotron emission of cosmic-ray electrons, and $21\cm$ line 
emission from \HImin. 
Attempts to separate the EGRB from the Galactic foreground have used either 
two-dimensional sky maps of these tracers, manipulated at various degrees 
of sophistication, or three-dimensional models of the Galactic components, 
constructed from such maps. 

Kraushaar et al. (1972) have assumed that the Galactic \gama-ray emission 
is proportional to the column density of \HI inferred from $21 \cm$ line 
observations. 
Analyzing the data of the OSO-3 satellite under this assumption, they found a 
residual isotropic background, generally softer than the Galactic emission, 
with intensity $I_x=3.0\pm0.9$ above $100\MeV$, in units of $10^{-5} \Iunits$. 
A similar analysis of the SAS-2 data resulted in a lower intensity, 
$I_x\simeq 1.0\pm0.4$, with a spectral index 
$p\simeq2.7_{-0.3}^{+0.4}$ \cite{Fichtel78}.  
Osborne, Wolfendale \& Zhang (1994) have analyzed the early data of the 
\emph{Energetic Gamma-Ray Experiment Telescope} (EGRET), 
by removing a component correlated with a $21\cm$ line survey and a residual 
latitude-dependent component, finding $I_x=1.10\pm0.05$ and $p=2.11\pm0.05$. 
In the most recent analysis of the EGRET data \cite{Sreekumar98}, an elaborate 
three-dimensional Galactic model \cite{Bertsch93,Hunter97} was employed, 
giving $I_x=1.45\pm0.05$ and $p\simeq2.10\pm0.03$. 
These examples demonstrate the sensitivity of the results to the Galactic 
model used. 

In this paper we analyze the high energy \gama-ray data set of EGRET 
--- which has the highest resolution and the lowest instrumental background 
presently available --- 
in search for robust, model independent constraints on the EGRB flux. 
We begin by presenting the data used in the analysis in \S \ref{sec:data}. 
We discuss both the EGRET \gama-ray data analyzed and the Galactic tracers, 
used in parts of the analysis to identify the Galactic foreground. 
The preparation of the data for the analysis (removal of various sources of 
noise, smoothing, binning and error estimation) is described. 

The Galactic \gama-ray foreground is minimal towards the Galactic poles. 
Hence, we begin our analysis in \S \ref{sec:galactic_poles} by examining 
the polar \gama-ray intensity. 
We show that the average EGRET-measured intensity near the poles is low, 
$I_{pole}=1.20\pm0.08$ in latitudes $|b|> 86\de$. 
This value is similar to previous estimates of the EGRB flux (and lower than 
the most recent estimate), even before Galactic foreground subtraction. 
We briefly estimate the Galactic contribution to the polar \gama-ray 
intensity, postponing a detailed discussion of the Galactic model to Appendix 
\ref{sec:galactic_model}, with a conservative estimate $I_{gal}\simeq 0.6-1.2$ 
and a lower limit of $I_{min}\simeq 0.4$. 
These estimates yield $I_x\simeq 0-0.6$, with a robust upper limit of 
$I_x\la 0.8$, possibly attributing the entire measured polar flux to 
Galactic foreground. 

Next, we study the high-latitude ($|b|>42\de$) EGRET data in 
\S\ref{sec:high_latitude_analysis}, 
finding an average latitude profile that is disk-like ($\sim 1/\sin|b|$) 
for $|b|\la70\de$ and steeper for $|b|\ga70\de$. 
This indicates that the Galactic foreground dominates the sky at all 
latitudes. 
We show that the high-latitude \gama-ray profile is very well fit by a 
model-motivated combination of two Galactic tracers: 
high frequency ($23\GHz$) synchrotron emission, and \HI column density 
inferred from $21\cm$ line emission. 
This fit is \emph{not} improved by including an isotropic component, 
indicating a small extragalactic contribution: $I_x<0.5\,(\at99\CL)$. 
The observed steep high latitude \gama-ray profile is inconsistent with 
isotropy. 
Attempts to reproduce this profile as a combination of any Galactic (steep) 
tracer and an isotropic (flat) profile yield similar upper limits, where 
most tracers favor a much lower isotropic flux. 
  
In \S \ref{sec:all_sky_analysis} we perform an all-sky analysis of the 
EGRET data, using correlations between the \gama-ray data and various Galactic 
tracers in order to identify and to subtract the Galactic foreground. 
Previous studies which have employed such methods in order to measure the EGRB 
are reviewed, and their results are recovered by making the same assumptions. 
We demonstrate the high sensitivity of the results of such correlation-based 
techniques to the choice of Galactic tracer, the number of tracers used and 
the part of the sky examined. 
We conclude that the Galactic foreground is not linear in the Galactic 
tracers, reflecting the interactions between different components of the 
Galaxy, as well as local features. 
Consequently, correlation-based methods can not be used to reliably determine 
the EGRB flux, but, rather to impose loose upper limits on it. 
We thus find a robust upper limit, $I_x \mla 1.0$, but present evidence for a 
much lower EGRB flux. 
For example, if the Galactic \gama-ray emission is assumed to follow the large 
scale structure of the Galaxy, we find $I_x\la0.6$.  

The most recent study of the EGRB \cite{Sreekumar98}, based on the EGRET data 
and an elaborate Galactic model, is reviewed and analyzed in 
\S \ref{sec:Sreekumar98}. 
We show that at least part of the reported EGRB flux must be associated with 
Galactic foreground. 

In Appendix \ref{sec:galactic_model} we describe the Galactic model used in 
\S \ref{sec:galactic_poles} to estimate the \gama-ray foreground towards the 
Galactic poles, and discuss its uncertainties and underlying assumptions. 
At the present, only crude estimates can be derived for the Galactic polar \gama-ray 
foreground, with insufficient accuracy for an identification of the weak 
\gama-ray background: 
Galactic emission could easily account, within present uncertainties, for the 
entire \gama-ray sky. 

We summarize our results in \S\ref{sec:discussion}, describing the emerging 
picture regarding the EGRB and the prospects for conclusively identifying it 
in the future. 
We present the implication of our results for the Galactic model of \gama-ray 
emission and the associated Galactic components, 
and for extragalactic systems involving gamma-ray emission, such as 
quasar jets (blazars), intergalactic shocks and ultra-high energy cosmic rays.

\section{Data}
\label{sec:data}

\subsection{Gamma-ray Data}
\label{subsec:gamma_ray_data}

We study the high-energy \gama-ray data measured by the Energetic Gamma-Ray 
Experiment Telescope (EGRET) aboard the Compton Gamma-ray Observatory (CGRO), 
providing the highest resolution and the lowest background data currently 
available. 

We have mainly used an all-sky EGRET map above $100\MeV$ from NASA's SkyView 
facility \cite{SkyView},
including data from 1991 April 22 to 1995 October 3 (EGRET cycles 1-4). 
We do not use the data in other energy bins, because the spectral data is 
shown to be of little use for our purpose, mainly because the EGRB is found 
to be much weaker than the Galactic foreground and because a-priori information 
regarding the EGRB spectrum is poor (see \S \ref{sec:discussion}). 
For brevity, the \gama-ray intensities discussed below are for energies 
above $100\MeV$, and in units of $10^{-5}\Iunits$, unless stated otherwise. 

The data was sampled into an all sky map with a $2\de$ resolution, using 
triangular decomposition. 
The point sources found in this data, summarized in the third EGRET catalog 
of high energy \gama-ray sources \cite{Hartman99}, were removed from the map. 
In some stages of the analysis, in order to correct for the limited angular 
resolution of EGRET, the data was smoothed with a two-dimensional Gaussian 
filter of standard deviation $4\de$, limited to $30\de$ in extent. 
This reproduces the main features of the calibrated instrument point-spread 
function (PSF) above $100\MeV$: $67\%$ containment within $5\de.85$ and $50\%$ 
containment within $4\de.5$ \cite{Thompson93}. 
Our statistical analysis always involved binning of the data at a minimal 
angular separation of $4\de$, such that the smoothing procedure had only a 
negligible effect on our results. 

The EGRET exposure during the measurement period (cycles 1-4) varied by 
roughly an order of magnitude across the sky, in the range 
$(2-22)\times10^8\cm^2\se$ \cite{Hartman99}. 
This varying exposure should be taken into account when evaluating the 
statistical noise of the \gama-ray data. 
For simplicity, when performing an all sky analysis 
(\S \ref{sec:all_sky_analysis}), a conservative value for 
the effective exposure has been chosen as $6\times10^8\cm^2\se$. 
We have verified that a change in this value bears a negligible effect on our 
results. 

The \gama-ray intensity measured in the direction of the Galactic plane and 
towards the inner Galaxy was found to be highly dominated by Galactic 
foreground. 
It is difficult to model the Galactic emission in these regions at sufficient 
accuracy to identify the far weaker extragalactic background.  
Hence, following Sreekumar et al. (1998), we exclude the Galactic plane 
($|b|<10\de$) and the inner Galaxy ($|b|<30\de$, $|l|<40\de$) from our 
analysis. 
An all-sky map of the analyzed \gama-ray data is presented in 
Figure \ref{fig:sky_maps} (upper left panel).

\subsection{Galactic tracers}
\label{subsec:galactic_tracers_data}

Attempts to separate the EGRB from the strong Galactic \gama-ray foreground 
have often modeled the \gama-ray emission of the Galaxy using Galactic 
tracers: observations of the sky at various frequencies, featuring different 
components of the Galaxy. 
Galactic \gama-ray emission is believed to arise from the interactions of 
Galactic CRs with gas and radiation. 
Hence, the Galactic components relevant for its \gama-ray emission are 
CR electrons and protons, gas (mostly \HImin, \HII and \HHmin) 
and radiation fields (mainly in the optical and in the infrared bands). 

The distribution of CR electrons in the Galaxy may be deduced from maps of 
radio synchrotron radiation, emitted as these electrons gyrate in interstellar 
magnetic fields, although this requires some assumptions regarding the 
strength and the distribution of magnetic fields through the Galaxy. 
Information regarding the \emph{local} electron flux is obtained by direct 
measurements carried out above $10\GeV$, where the solar modulation has little 
effect, at the top of the Earth atmosphere.  
The CR electron spectrum may be inferred partially from radio and soft 
\gama-ray observations, complemented by direct measurements above $10\GeV$ 
\cite{Longair81,DuVernois01,Casadei03}. 
The CR proton distribution may be estimated with some assumption regarding the 
CR proton-to-electron ratio, usually presumed constant throughout the Galaxy. 
The local flux and spectrum of CR protons and nuclei are measured directly as 
well, at the top of the atmosphere, for energies above $1\GeV$ per nucleon. 

The distribution of gas in the Galaxy may be deduced from various forms of 
radiation emitted by the gas or from its correlations with other Galactic 
components, such as dust \cite{Puget76}. 
The distribution of \HI may be inferred from maps of $21\cm$ line emission and 
absorption, optical $\mbox{H}\alpha$ ($3\rightarrow 2$ transition) emission 
\cite{Finkbeiner03}, and UV Lyman-series measured in absorption against 
background stars \cite{Dickey90}. 
The \HH distribution may be derived from the lowest frequency transition 
of \CO at $115\GHz$, which is well correlated with the molecular hydrogen 
to within a factor of $2$ when integrated over large regions \cite{Dame01}. 
The model for the \HII distribution is based on propagation effects 
of radiation through the ISM, most importantly the dispersion, temporal and 
angular broadening, and scintillation of pulsar radio signals \cite{Cordes02}. 

Three important radiation fields are Compton-scattered by CR electrons, thus 
contributing to the Galactic \gama-ray emission: 
the Galactic infrared (IR) background, the Galactic optical background, and 
the cosmic microwave background (CMB). 
The IR and optical distributions through the Galaxy are uncertain, estimated 
by integrating the IR and optical Galactic emissivities inferred from models 
and observations (see Appendix \ref{sec:galactic_model}). 

In some parts of our analysis, we model the Galactic \gama-ray emission using 
the Galactic tracers mentioned above. 
We thus study seven Galactic tracers (see Table \ref{tab:Tracers}): 
\begin{enumerate}
\item{A $408\MHz$ survey \cite{Haslam82}, consisting mostly of synchrotron 
radiation. }
\item{A $23\GHz$ map found as synchrotron foreground for WMAP 
\cite{Bennett03}.}
\item{An \HI column density map based on $21\cm$ line emission \cite{Dickey90}}
\item{An \HI survey based on the \Halpha ($3\rightarrow2$) transition 
\cite{Finkbeiner03}. }
\item{A column density map of dust based on IR measurements \cite{Schlegel98}.}
\item{A synthetic \HII column density map from a pulsar-based model 
\cite{Cordes02}.}
\item{A synthetic map of Galactic \emph{large-scale} structure, based on 
this pulsar model.}
\end{enumerate}

\begin{figure}
\epsscale{1.0}
\plottwo{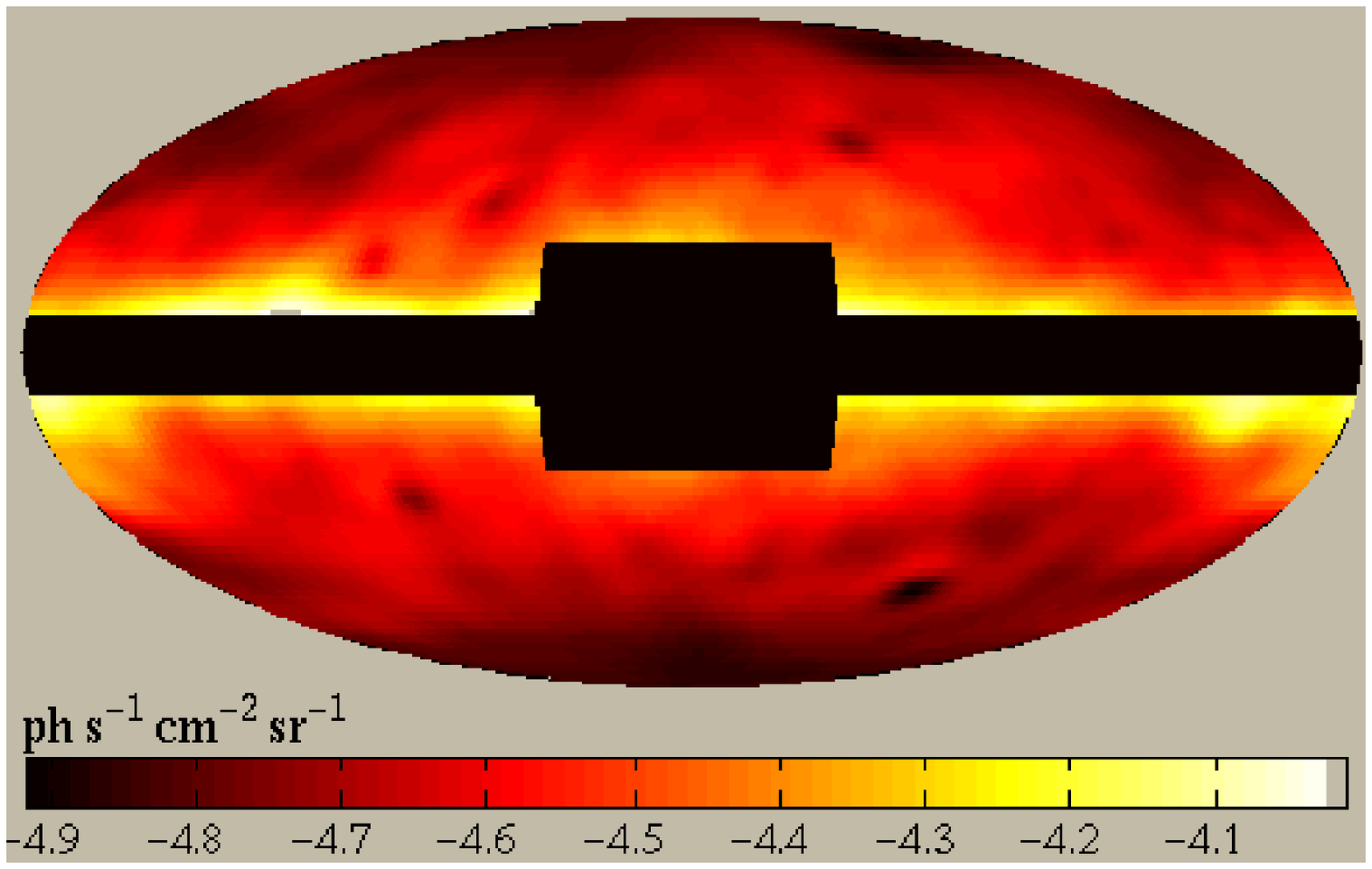}{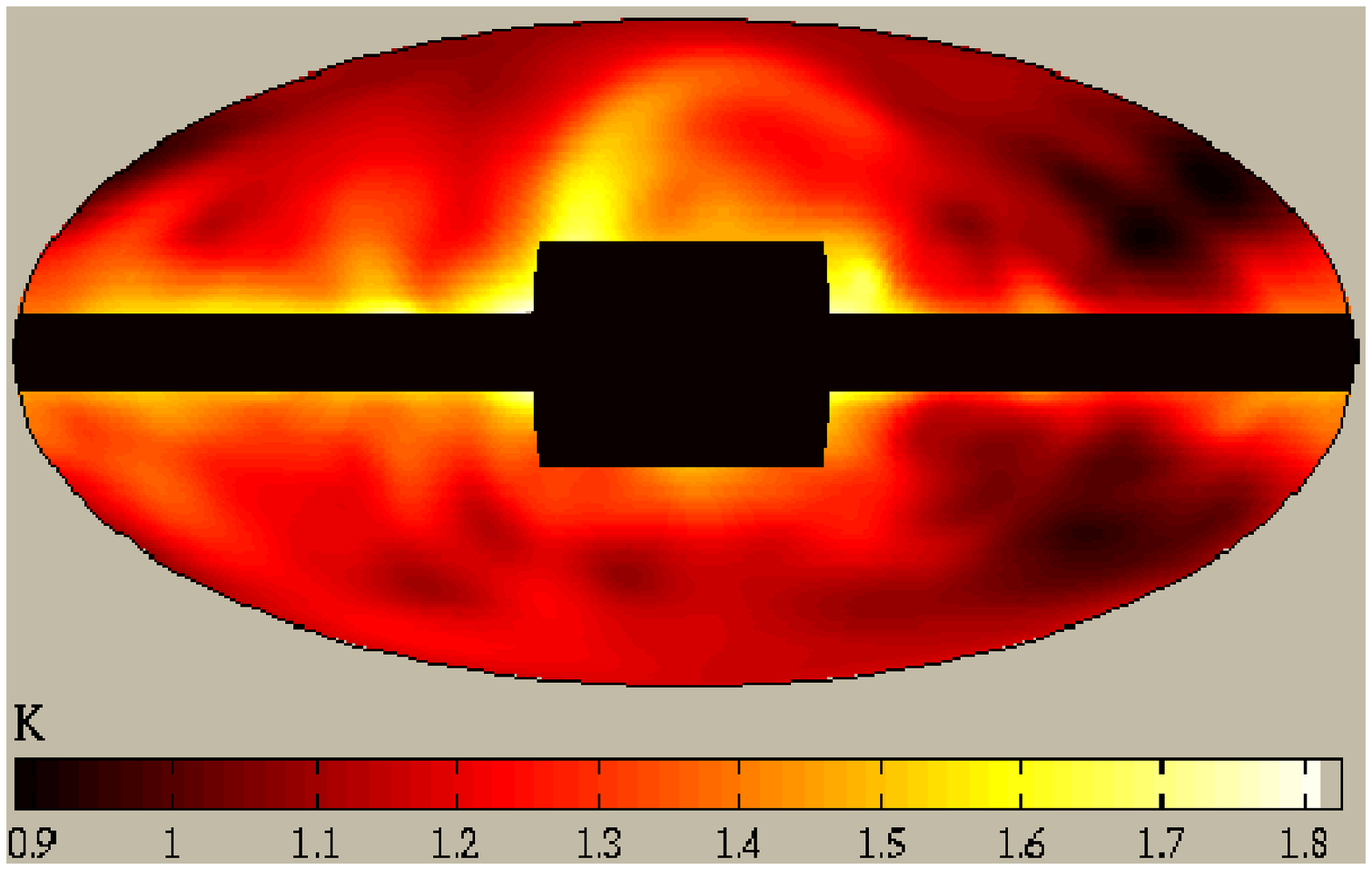}
\epsscale{2.25}
\plottwo{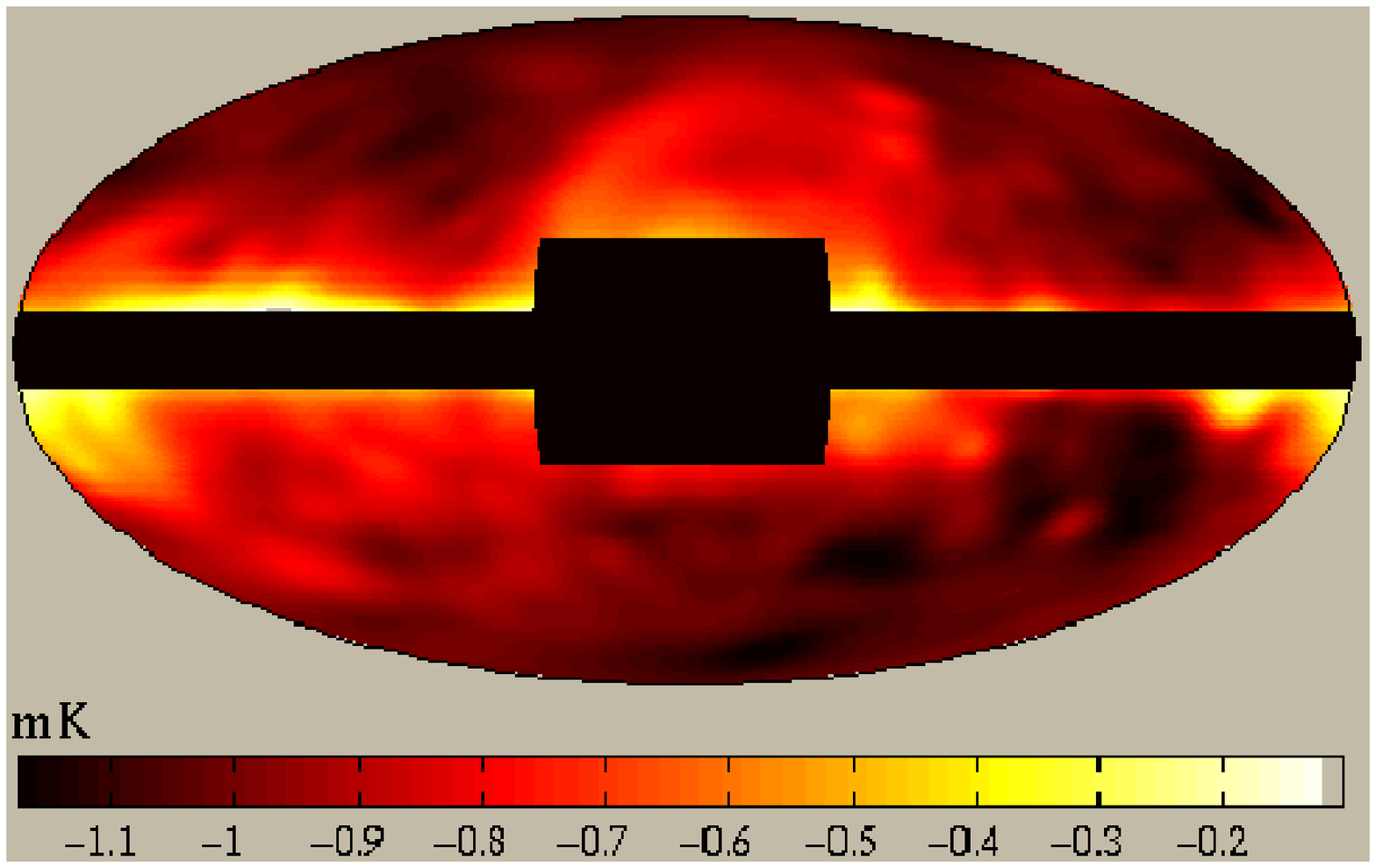}{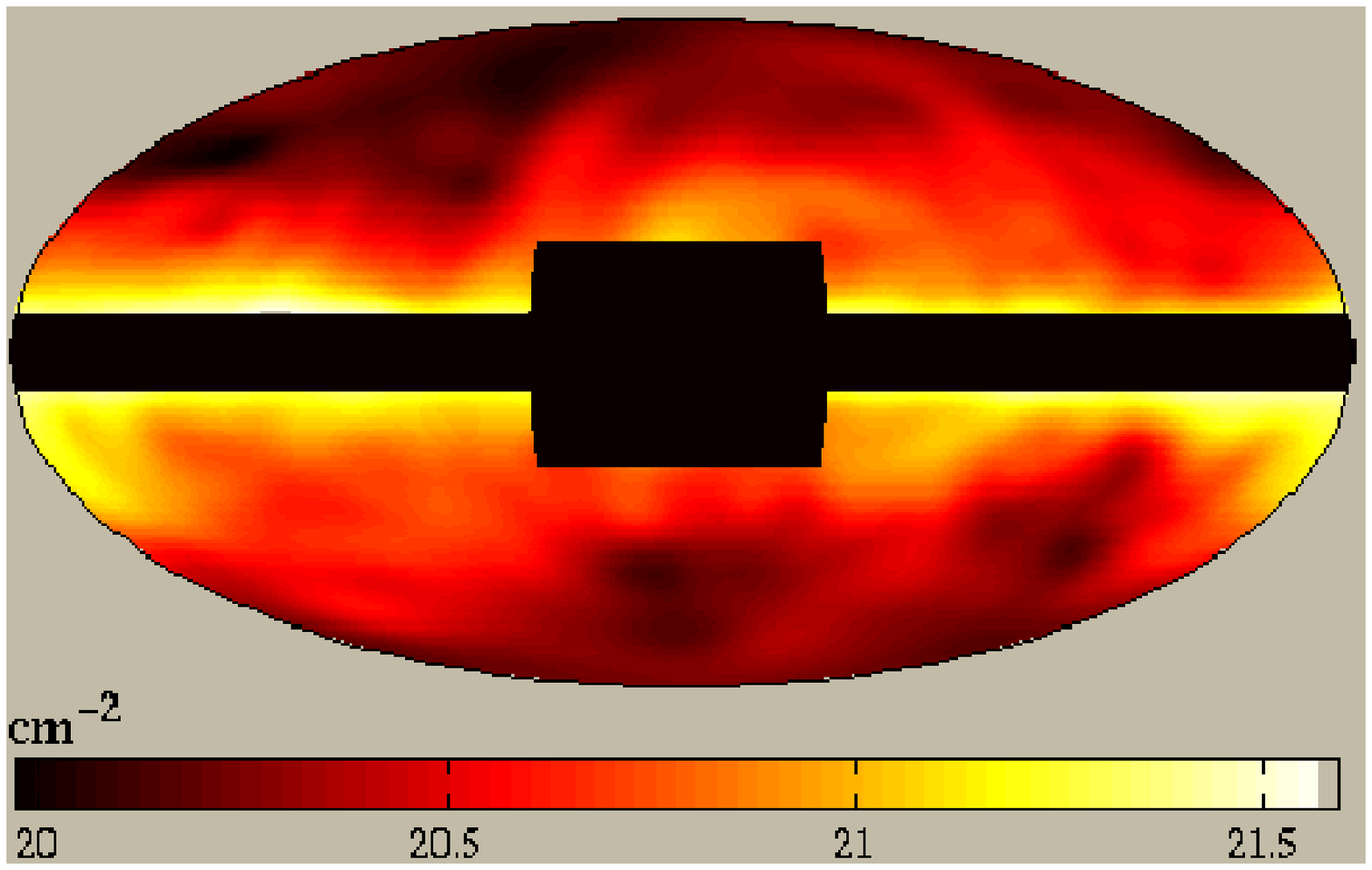}
\epsscale{5.06}
\plottwo{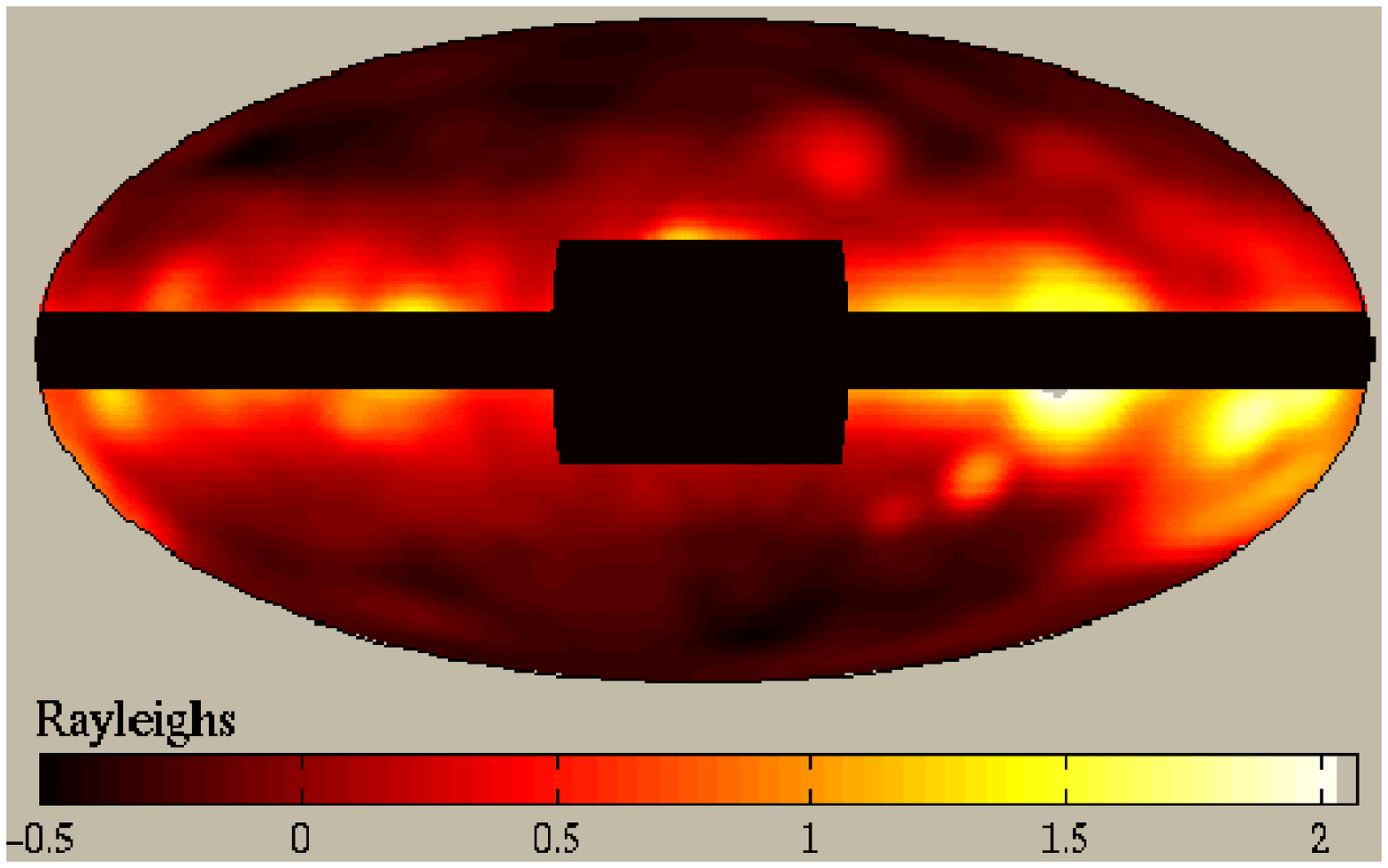}{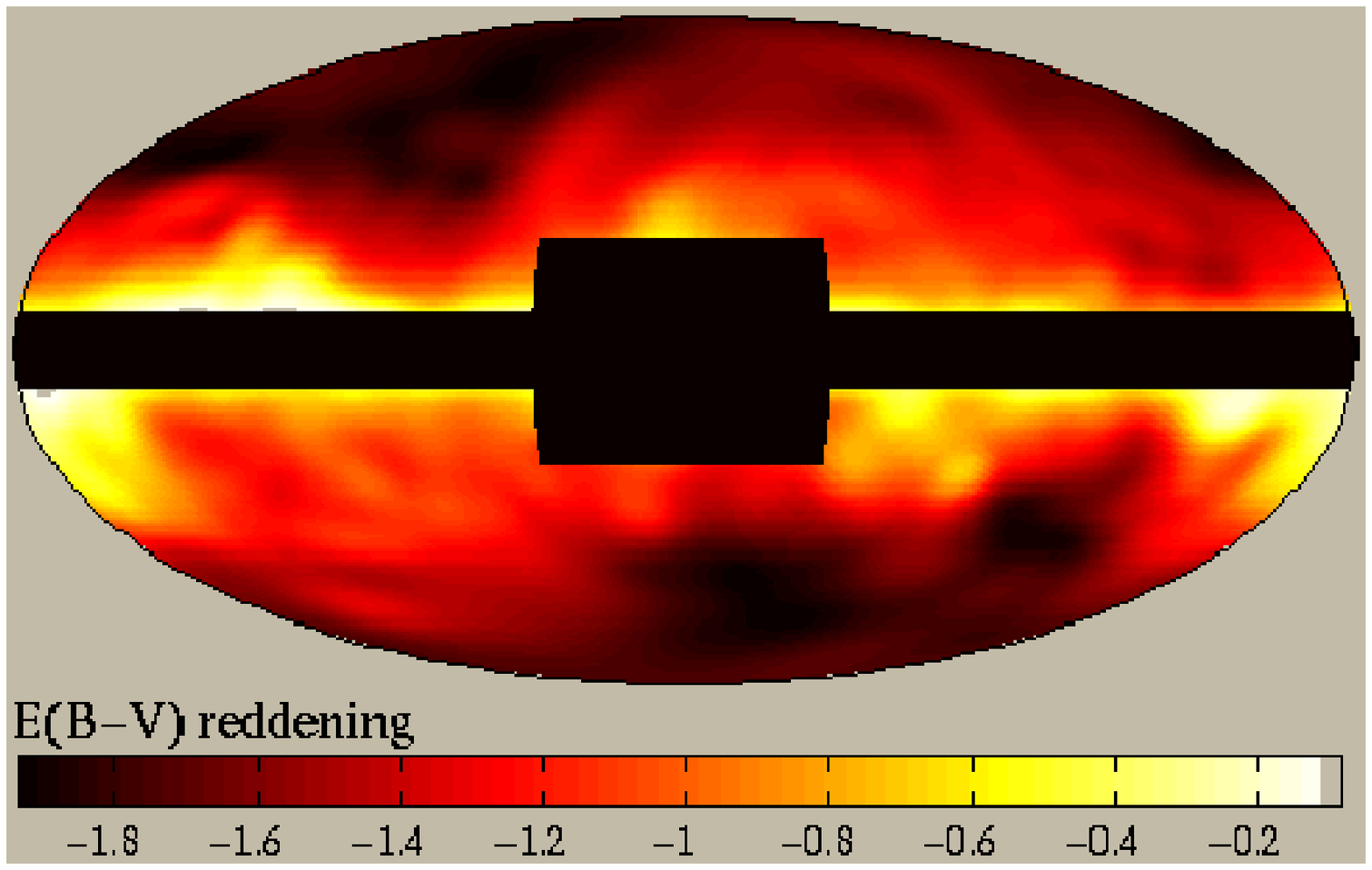}
\caption{ 
Hammer-Aitoff projection of the high energy \gama-ray EGRET all-sky map 
(\emph{upper left}) and of some Galactic tracers: 
the $408\MHz$ synchrotron map (tracer 1, \emph{upper right}), 
the $23\GHz$ synchrotron map (tracer 2, \emph{middle left}), 
the $21\cm$ map (tracer 3, \emph{middle right}), 
the \Halpha map (tracer 4, \emph{bottom left}), 
and the dust map (tracer 5, \emph{bottom right}). 
All maps have been smoothed with a filter function designed to imitate the 
EGRET point spread function above $100\MeV$ (\S \ref{sec:data}). 
The Galactic plane and the inner Galaxy were excluded because their signal 
is very high. 
Point sources were removed from the \gama-ray and from the $408\MHz$ data. 
In addition, the contributions of the CMB and of an estimated extragalactic 
component were removed from the $408\MHz$ map. 
Color scales are logarithmic with a base $10$.
}
\label{fig:sky_maps}
\end{figure}

In order to compare the \gama-ray data with the Galactic tracers directly, 
all-sky maps of these tracers were convolved with the same Gaussian filter 
applied to the EGRET data, before being analyzed.
For the same reason, the Galactic plane ($|b|<10\de$) and the inner Galaxy 
($|b|<30$, $|l|<40\de$) were excluded from the tracer maps. 
The resulting maps of the empirical, non-synthetic tracers (tracers 1-5), 
depicting the regions relevant to our analysis, are shown in 
Figure \ref{fig:sky_maps}, alongside the similarly prepared EGRET map. 

We use tracers of Galactic synchrotron emission in two different frequencies, 
$408\MHz$ and $23\GHz$, because the spectral indices of this radiation vary 
significantly across the sky, resulting in different map characteristics at 
different frequencies \cite{Bennett03}. 
We use the dust map because of its correlation with the gas, as discussed 
above. 
We do not use an \HH Galactic tracer, e.g. as inferred from the $115\GHz$ 
rotational transition of \COmin, because the molecular gas is mostly found 
in low ($b\la10\de$) latitudes \cite{Dame01}, excluded from the analysis. 

\begin{figure}
\epsscale{1.0}
\plotone{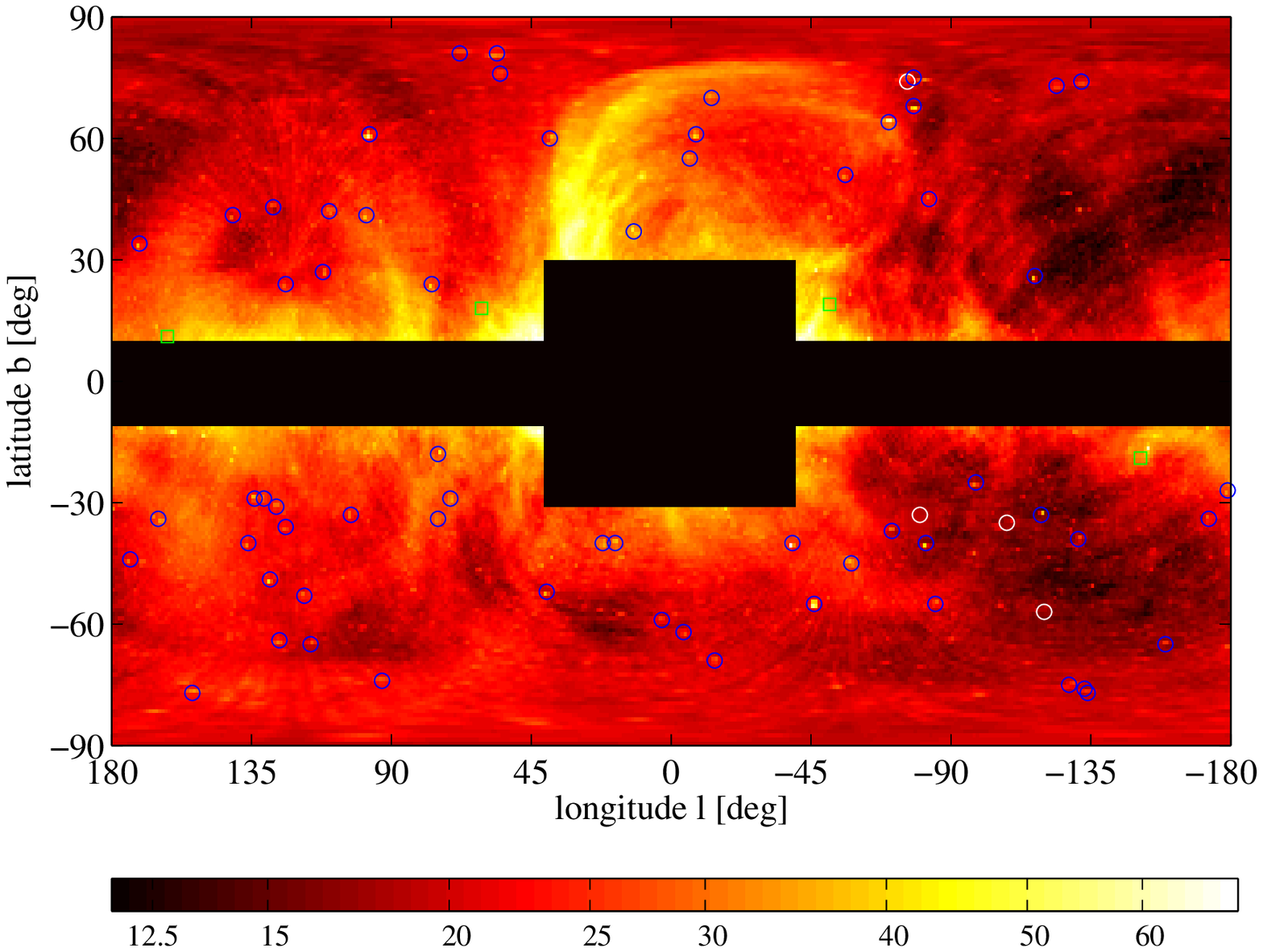}
\caption{ 
The radio sky at $408\MHz$ in Galactic coordinates \cite{Haslam82} in a 
rectangular projection. 
The data from low latitudes ($|b|<10\de$) and from the Galactic central region 
($|b|<30\de$,$|l|<40\de$) have been blocked, for comparison with the \gama-ray 
map of Sreekumar et al. (1998) and in order to enhance the contrast of the 
image. 
Point sources identified by our algorithm are shown (blue circles). 
Four bright point sources (white circles) and four localized regions of bright 
emission associated with extended Galactic structure (green squares) were 
removed from the image using linear interpolation, in order to enhance the 
contrast, although they are maintained in the analysis.  
Color scale: effective temperature in K.
}
\label{fig:map_408_MHz}
\end{figure}

The $408\MHz$ map includes, in addition to the diffuse Galactic synchrotron 
radiation, small contributions from free-free emission within the Galaxy 
(Dickinson, Davies \& Davis 2003), the CMB, radio point sources, and an 
unknown extragalactic radio background (ERB). 
We remove the contributions of the CMB and the ERB by modeling their combined 
contribution to the map as an average $6\K$ enhancement in antenna 
temperature, considered smooth over the relevant $\gg 1\de$ scales 
\cite{Longair72,Lawson87}. 
Unresolved point sources and bright resolved sources not related to Galactic 
large-scale structure were removed from the map using linear interpolation. 
Figure~\ref{fig:map_408_MHz} shows the $408\MHz$ all-sky map, highlighting 
the various point sources. 
The $23\GHz$ synchrotron map was deduces from the first year observations of 
WMAP, using the maximum entropy method (MEM) with the $408\MHz$ map as a prior 
\cite{Bennett03}. 
Although the residuals of this map are small ($<1\%$), we attribute to this 
tracer a minimal systematic error of $5\%$, on all scales. 

The synthetic \HII map was calculated using the NE2001 code, simulating a 
model based on the propagation effects of pulsar radiation through the ISM 
\cite{Cordes02}. 
This is a synthetic map, relying on some simplifying assumptions instead of 
being inferred directly from observations, and is thus treated here with some 
caution. 
The NE2001 model calculates the emission from several global and local 
Galactic components.  
The global components describe the large-scale structure of the Galaxy, 
including a thick Galactic disk of scale height $1\kpc$ and scale radius 
$20\kpc$, an inner thin disk of scale height $140\pc$ and scale radius 
$11\kpc$, spiral arms and a Galactic central bulge. 
The local components include low density regions near the sun such as bubbles, 
local objects such as the Vela supernova remnant, regions of intense 
scattering (clumps) and regions of very low density (voids).  
In addition to the \HII tracer (tracer 6), incorporating all the above 
components, we also produce a map of the sky based only on the global Galactic 
components outlined above. 
This synthetic tracer (tracer 7) corresponds to the large-scale structure of 
the Galaxy, and may be used, for example, as a first-order guess for the 
Galactic CR distribution.

\section{Galactic Poles}
\label{sec:galactic_poles}

Just as regions near the Galactic plane are ideal for studying the \gama-ray 
emission of the Galaxy, because they are dominated by it and exhibit various 
useful features, the Galactic polar regions are most suitable for identifying 
the extragalactic radiation, because of their 
lower Galactic foreground and fewer small-scale features. 
We begin with the assumption that the extragalactic \gama-ray background is 
approximately isotropic on large angular scales, say $\ga 5\de$. 
Thus, the average intensity in any $\ga 5\de \times 5\de$ patch of the 
\gama-ray sky imposes a robust upper limit on the EGRB flux, independent of 
any model for Galactic emission. 
In particular, the Galactic poles are the faintest regions in the \gama-ray 
sky, as well as in most Galactic tracers, and hence impose the tightest 
straightforward upper limits on the extragalactic component. 
Supplementing the \gama-ray intensities measured toward the poles with an 
estimate of the \emph{minimal} Galactic foreground, yields a more stringent, 
albeit slightly model-dependent, upper limit on the EGRB flux. 
We thus briefly present a calculation of the Galactic polar foreground, 
leaving a more detailed discussion to Appendix \ref{sec:galactic_model}.

\subsection{Measured Polar \gama-ray Intensity}
\label{subsec:polar_intensity}

The average \gama-ray intensity measured by EGRET within $10\de$ of the 
Galactic poles, as extracted from the raw \gama-ray sky map (after modeling 
point sources), is $I(|b|>80\de)=1.55\pm0.04$. 
However, the latitude dependence of the intensity near the Galactic poles, 
to be discussed in \S \ref{sec:high_latitude_analysis}, is strong: 
for latitudes $|b|\mga 70\de$ the intensity drops as the line of sight 
approaches the pole, faster than a disk-like $\sim 1/\sin|b|$ profile, 
almost linearly with latitude (see Figure \ref{fig:latitude_profiles}). 
This strong latitude dependence indicates that one should examine the 
intensity in the close vicinity of the poles in order to discover their 
true brightness. 
We thus examine the \gama-ray intensity measured closer to the poles; 
averaging the near-pole intensities, $I(b<-86\de)=1.17\pm0.11$ and 
$I(b>86\de)=1.23\pm0.12$, we obtain a better estimate of the polar intensity: 
\beq \label{eq:polar_intensity} I_{pole}=1.20\pm0.08 \fin \eeq 
Notice that no point sources have been identified in the close vicinity of 
the poles. 
Thus, the low polar intensities presented above can not be a result of poor 
modeling of the point sources. 
Since the latitude profile of the \gama-ray data is decreasing towards the 
poles, accounting for the small size of the polar caps examined with respect 
to the $67\%$ containment of the EGRET PSF would yield a polar intensity 
similar or even smaller than the value given in equation 
(\ref{eq:polar_intensity}).

The EGRET polar \gama-ray intensity determined above, $I_{pole}=1.20\pm0.08$, 
is similar to most estimates of the EGRB flux (and lower than the most recent 
estimate), even before taking into account the Galactic foreground.  
This conclusion alone imposes a robust, model-independent upper limit 
on the flux of any isotropic \gama-ray component such as the EGRB. 
However, both theoretical arguments (see Appendix \ref{sec:galactic_model}) 
and a phenomenological analysis of the high latitude EGRET data 
(see \S \ref{sec:high_latitude_analysis}), indicate that the Galactic 
contribution to the polar intensity is not negligible. 
On the contrary, these considerations suggest a substantial Galactic foreground towards the poles, permitting only an EGRB flux much lower than the above upper limit. In the following, we estimate the minimal level of Galactic foreground towards the poles, and discuss the implied maximal EGRB flux.

\subsection{Subtracting the Galactic Contribution}
\label{subsec:subtracting_galactic_contribution}

The Galactic \gama-ray foreground towards the poles can be crudely estimated 
from present models of the Galactic components involved in \gama-ray emitting 
processes. 
High energy ($>100\MeV$) \gama-ray emission from the Galaxy is dominated by 
three processes, providing roughly comparable contributions to the polar 
foreground at the Galactocentric position of the sun: 
bremsstrahlung of CR electrons in the interstellar gas, inverse-Compton 
scattering of optical and IR radiation fields by CR electrons, 
and nucleon-nucleon scattering of CR protons by the interstellar gas. 
The contributions of these processes depend on the distributions of gas, 
radiation and CRs, as well as on the overlap between these components along 
the line of sight. 
These distributions, in turn, may be inferred from the associated Galactic 
tracers (see Appendix \ref{sec:galactic_model}): the distribution of gas is inferred from its radiative consequences, the distribution of radiation fields is based on direct measurements combined with Galactic emissivity models, the CR electron distribution is deduced from radio synchrotron observations, and the CR proton distribution is assumed to follow the CR electrons, and its local flux is directly measured at the top of the Earth atmosphere. 
Using these tracers and making common assumptions regarding the average 
magnetic field at the Galactocentric position of the sun ($B\simeq 5\muG$), 
the CR scale height ($L\simeq 1\kpc$) and the overlap between the various Galactic components (uniform), the polar Galactic foreground is estimated to lie in the range (see Appendix \ref{sec:galactic_model} for details):
\beq \label{eq:galactic_estimate} I_{gal}\simeq0.6-1.2 \fin \eeq 

However, given the limited information we have regarding the CR distribution 
and the overlap between the relevant Galactic components along the line of 
sight, the true Galactic foreground towards the poles could lie outside the 
above estimated range. 
Nevertheless, upper limits on the Galactic magnetic field amplitude, the CR 
distribution scale height and the level of possible anti-correlations between 
the Galactic components, suggest that the Galactic foreground towards the 
poles can not be lower than $I_{min}\simeq 0.4$ 
(see Appendix \ref{sec:galactic_model}). 
This result is in accord with a previous calculations of the polar Galactic 
foreground by Fichtel et al. (1978), who found $I_{gal}=0.45-0.88$ but suggested that the Galactic foreground is stronger. 

Using the estimated polar intensity [equation (\ref{eq:polar_intensity})] and 
the minimal estimated Galactic foreground, $I_{min}=0.4$, we may impose 
a robust upper limit on the EGRB intensity, $I_x\la 0.8$. 
A more plausible estimate of the Galactic foreground 
[equation (\ref{eq:galactic_estimate})] yields a lower estimate:
\beq I_x\simeq 0-0.6 \fin \eeq
Similar results may be obtained without the above theoretical considerations, 
by using phenomenological arguments instead (e.g. the high-latitude analysis 
of \S \ref{sec:high_latitude_analysis}
or Galactic features evident in the \gama-ray data near the poles) to deduce 
that the Galactic contribution to the polar \gama-ray intensity is at least 
comparable to the extragalactic component. 
It is important to note that both our calculations 
(see Appendix \ref{sec:galactic_model}) and the phenomenological analysis 
(see \S \ref{sec:high_latitude_analysis}) indicate that the Galactic 
foreground towards the poles is \emph{higher} than $I_{min}$, 
suggesting a lower EGRB flux. 
In fact, the results of both methods are consistent with a negligible 
isotropic extragalactic component, or no such component at all. 
Finally note, that by considering a larger patch of the sky, $|b|>80\de$, 
one obtains a higher estimate of the EGRB flux: $I_x\simeq 0.3-0.9$. 
Although such an estimate probably overshoots the true EGRB flux, because the 
latitude profile of the \gama-ray data is steep at such latitudes, it is still lower than most previous estimates of the EGRB flux. 

To conclude, we find a low polar \gama-ray brightness, containing a 
significant and probably dominant, Galactic foreground. 
Hence, the \gama-ray intensity of the poles precludes an extragalactic 
isotropic \gama-ray background as strong as found by previous studies. 
The polar brightness may be used to impose a conservative upper limit on the 
EGRB flux, $I_x\la 0.8$, with minimal foreground subtraction.
However, theoretical and phenomenological evidence, presented in the following 
sections, suggests an EGRB flux much lower than this upper limit.

\section{High Latitude Analysis}
\label{sec:high_latitude_analysis}

After extracting information regarding the EGRB flux from the \gama-ray 
intensity measured towards the Galactic poles, and before analyzing the 
all-sky \gama-ray map, we study the EGRET data measured at high-latitudes, 
$|b|> 42\de$, where the Galactic foreground exhibits interesting features 
while remaining relatively low.  
We concentrate on the \emph{average latitude profile} of the data, exploiting 
its relative simplicity and insensitivity to local features, with respect to 
two-dimensional sky maps. 
Our analysis is largely phenomenological, imposing additional constraints on 
the EGRB flux that are robust and mostly model-independent. 

\begin{figure}
\epsscale{0.8}
\plotone{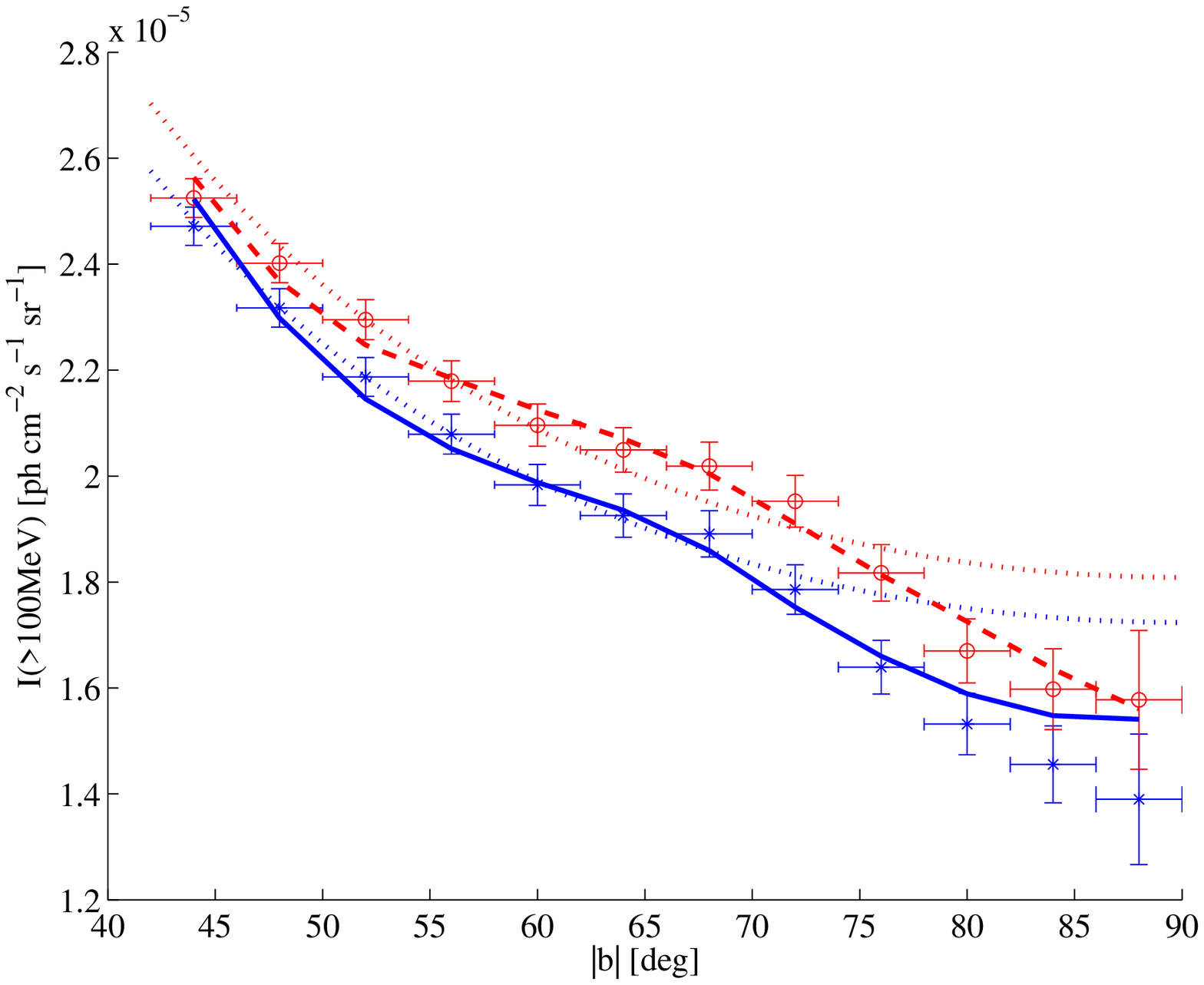}
\epsscale{1.1}
\plottwo{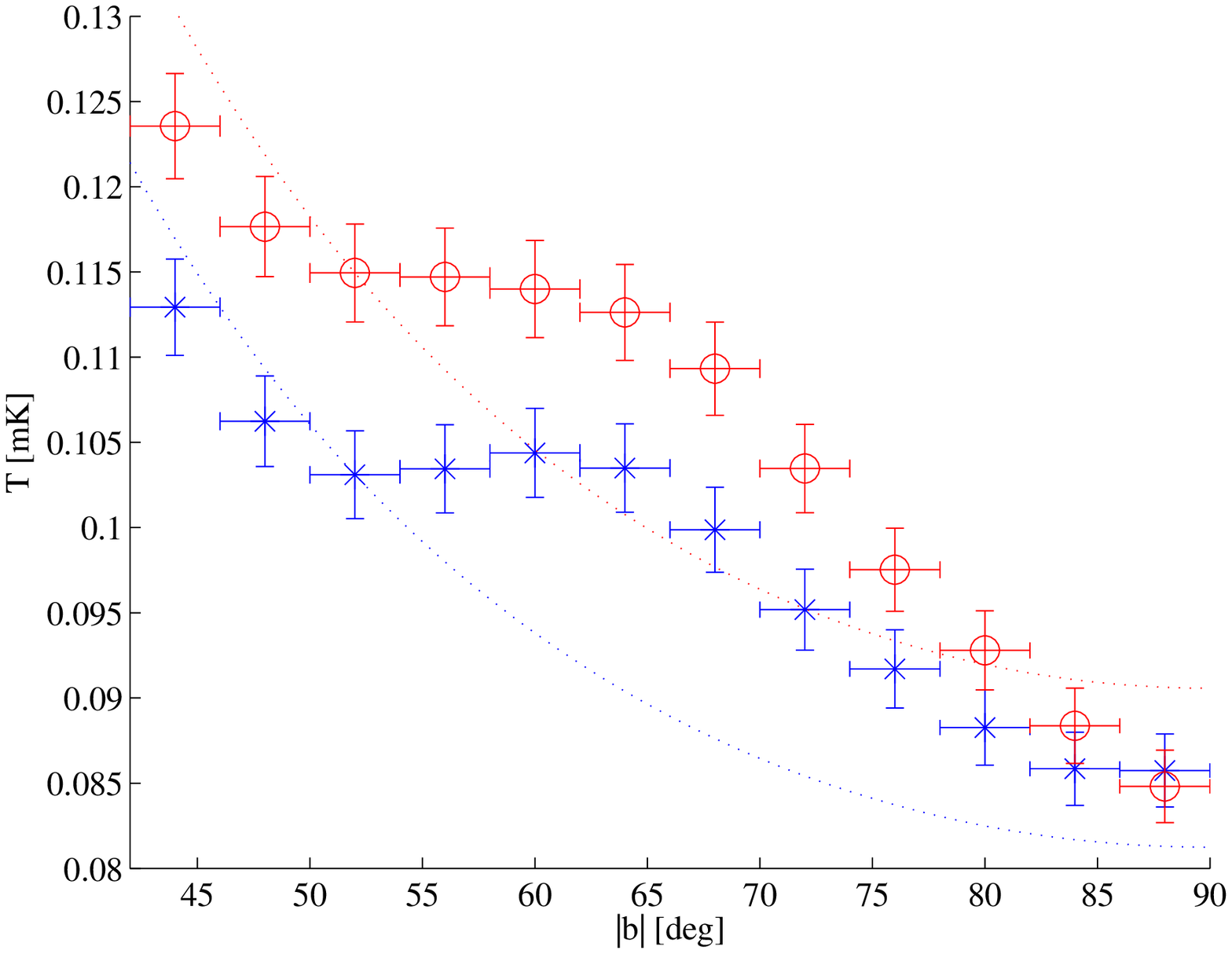}{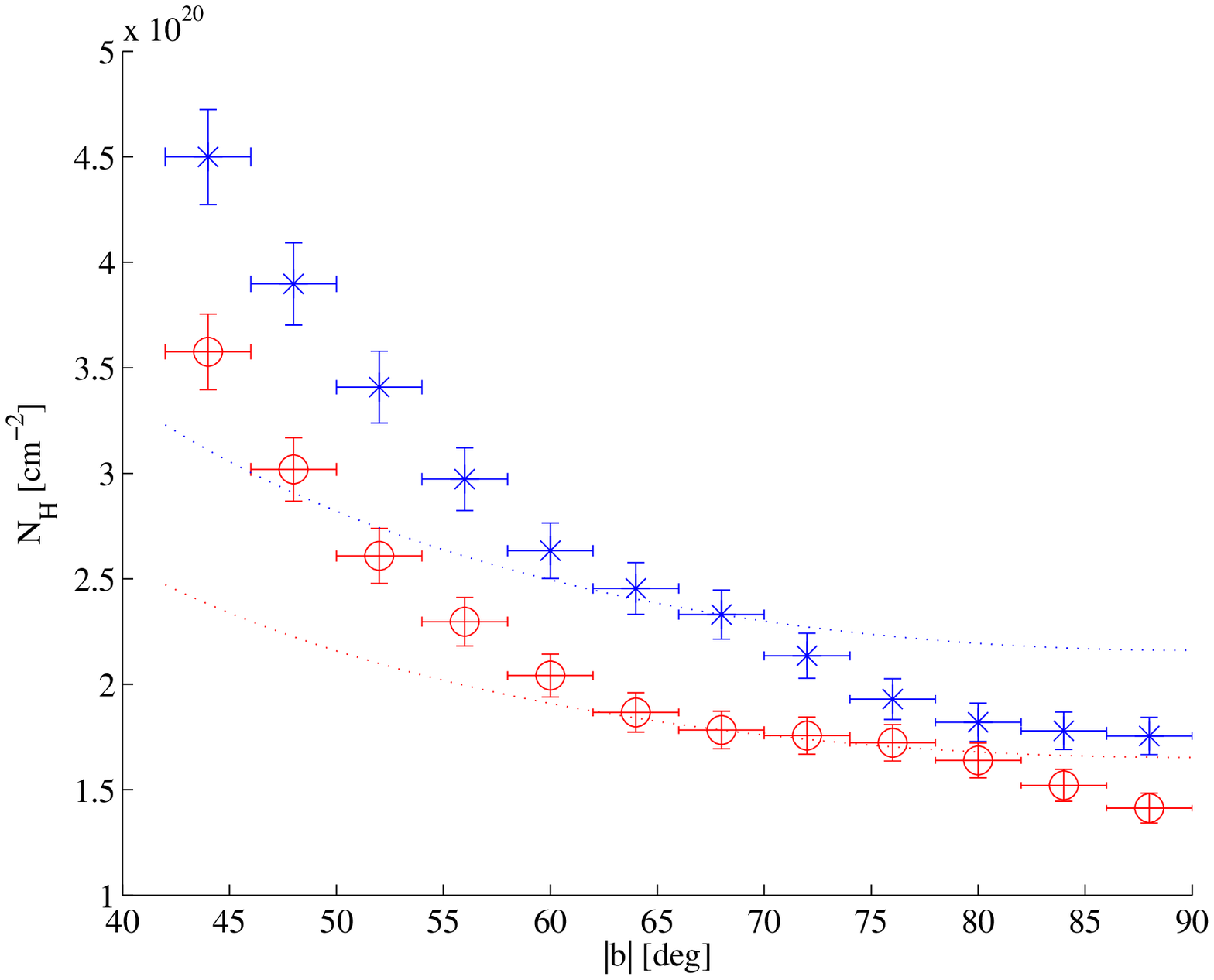}
\caption{ 
High latitude data of the EGRET \gama-ray intensity (\emph{upper panel}), 
effective temperature of synchrotron emission at $23\GHz$ (\emph{lower left}), and \HI column density inferred from $21\cm$ line emission 
(\emph{lower right}). 
Shown are the latitude profiles in the southern (\emph{stars}) and in the 
northern (\emph{circles}) Galactic hemispheres, along with characteristic 
$1/\sin|b|$ profiles (\emph{dotted lines}).
The \emph{upper panel} also presents linear combinations of the $23\GHz$ and 
the $21\cm$ data (see text) in the southern (solid line) and northern (dashed 
line) hemispheres. 
}
\label{fig:latitude_profiles}
\end{figure}

\begin{figure}
\epsscale{1.0}
\plottwo{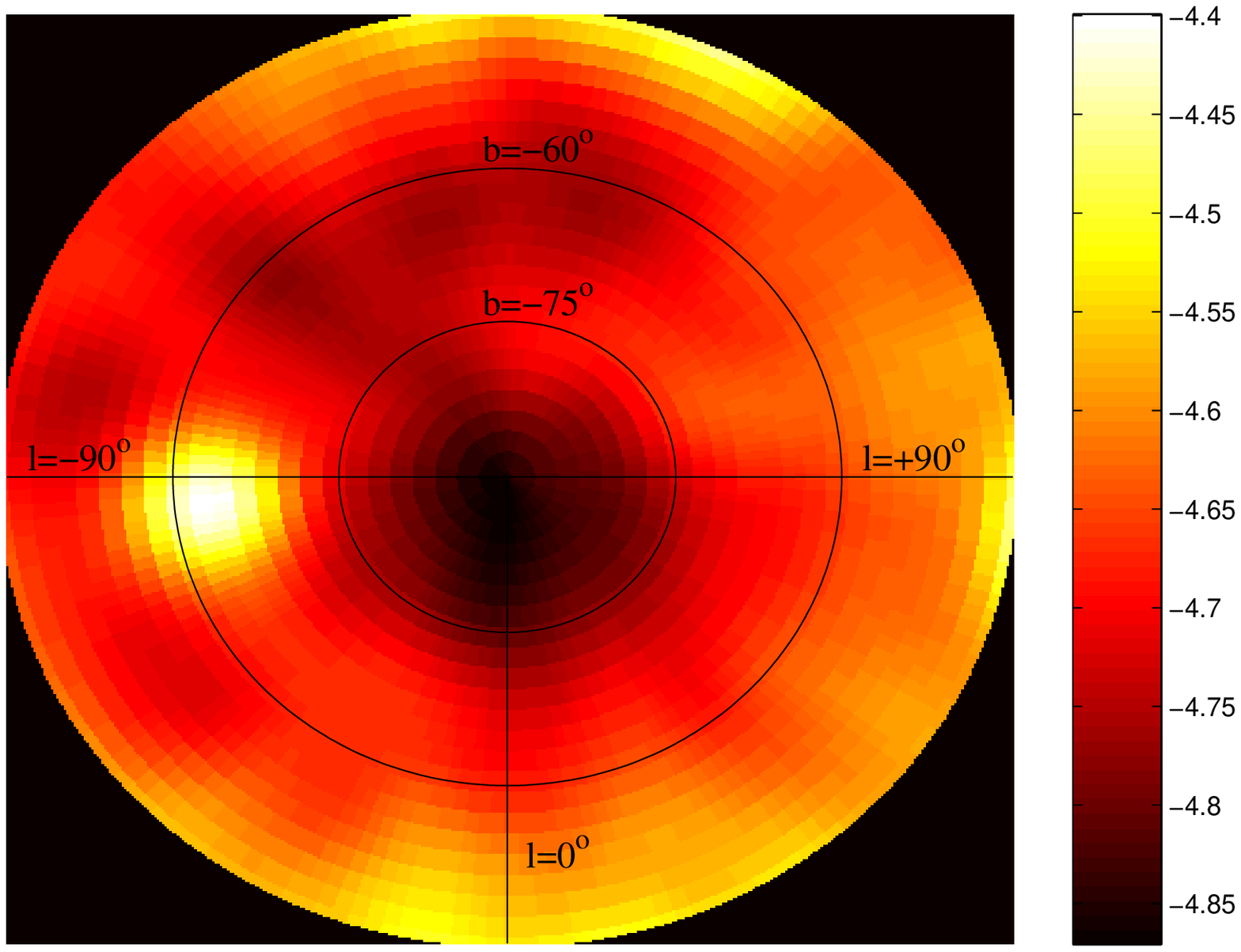}{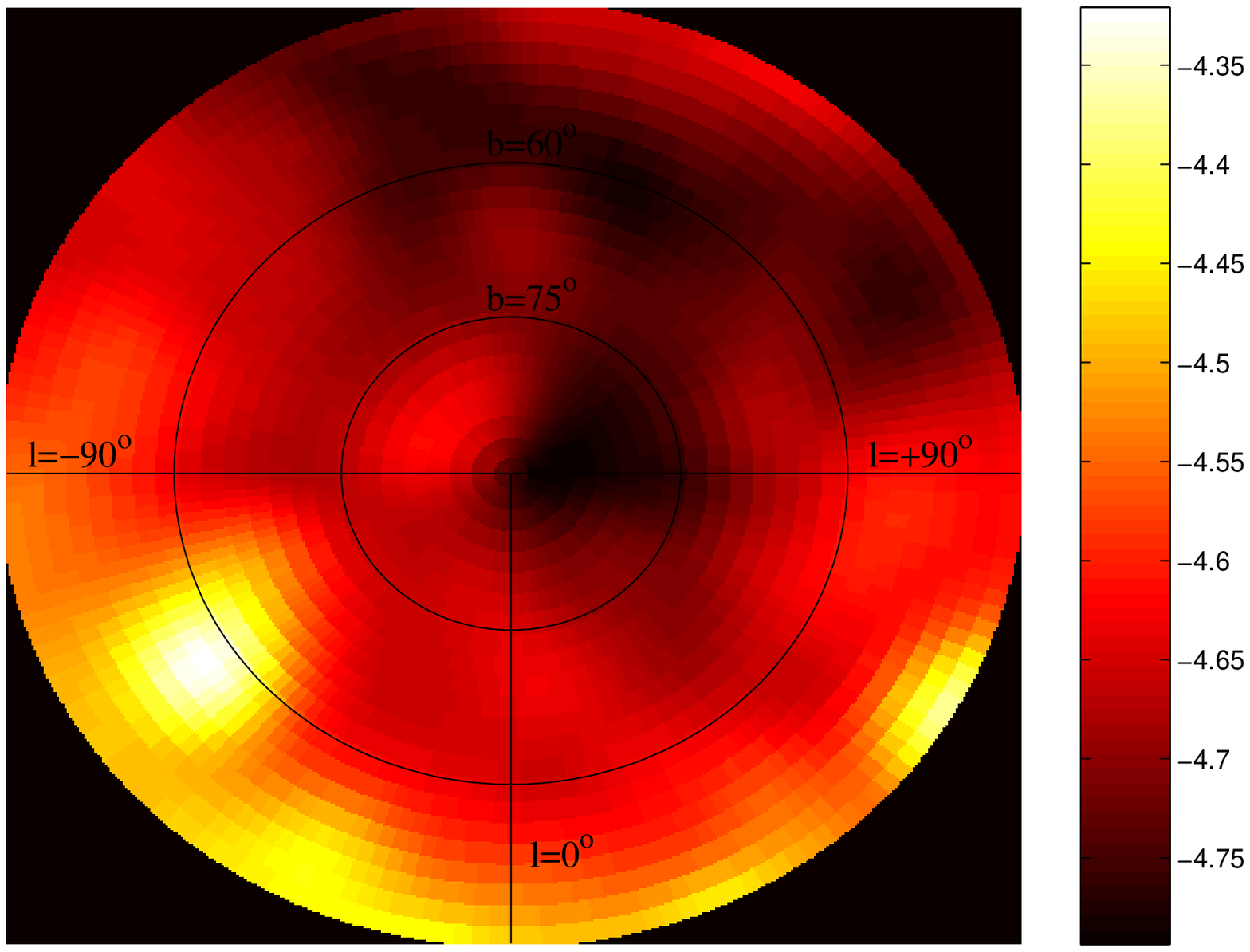}
\epsscale{2.25}
\plottwo{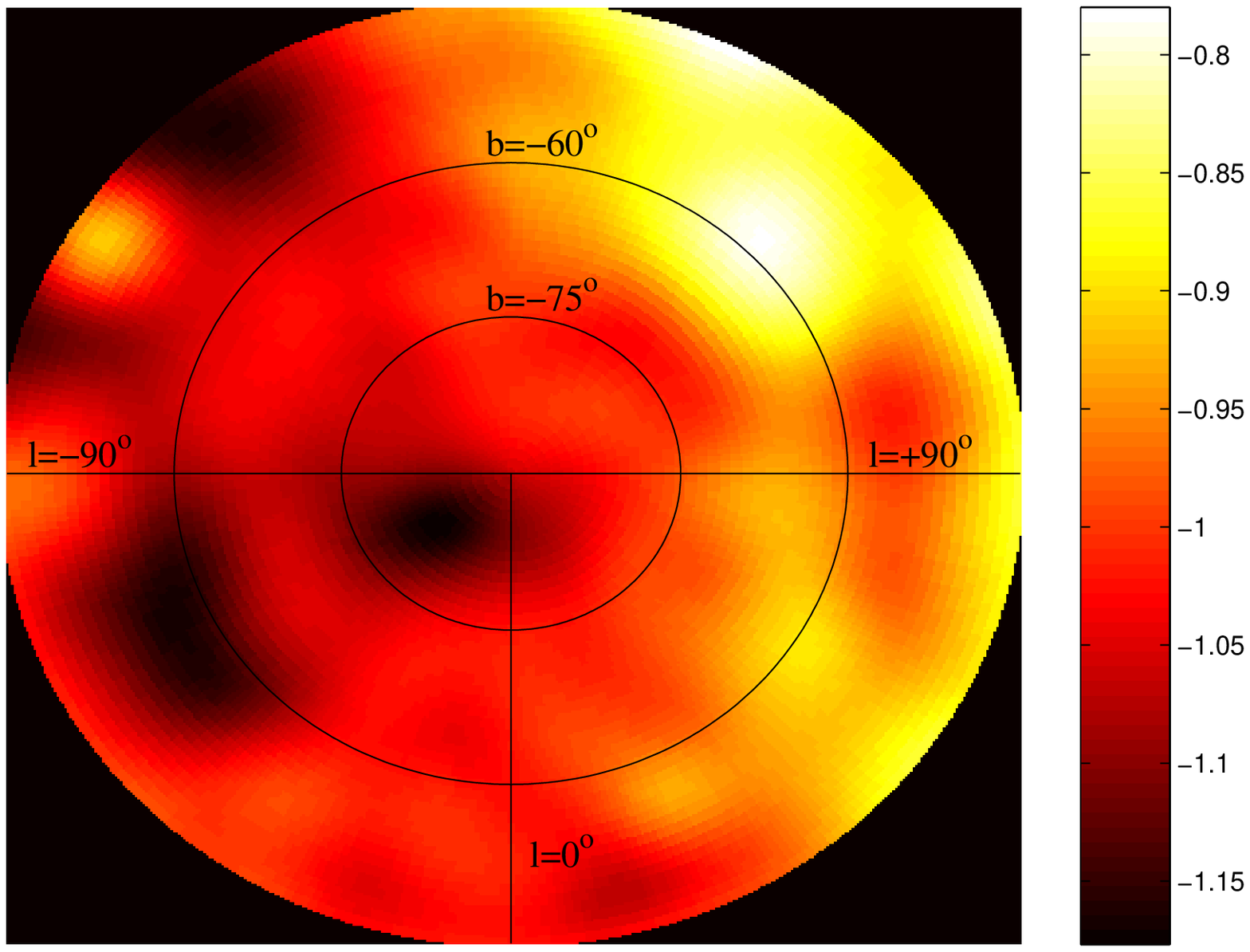}{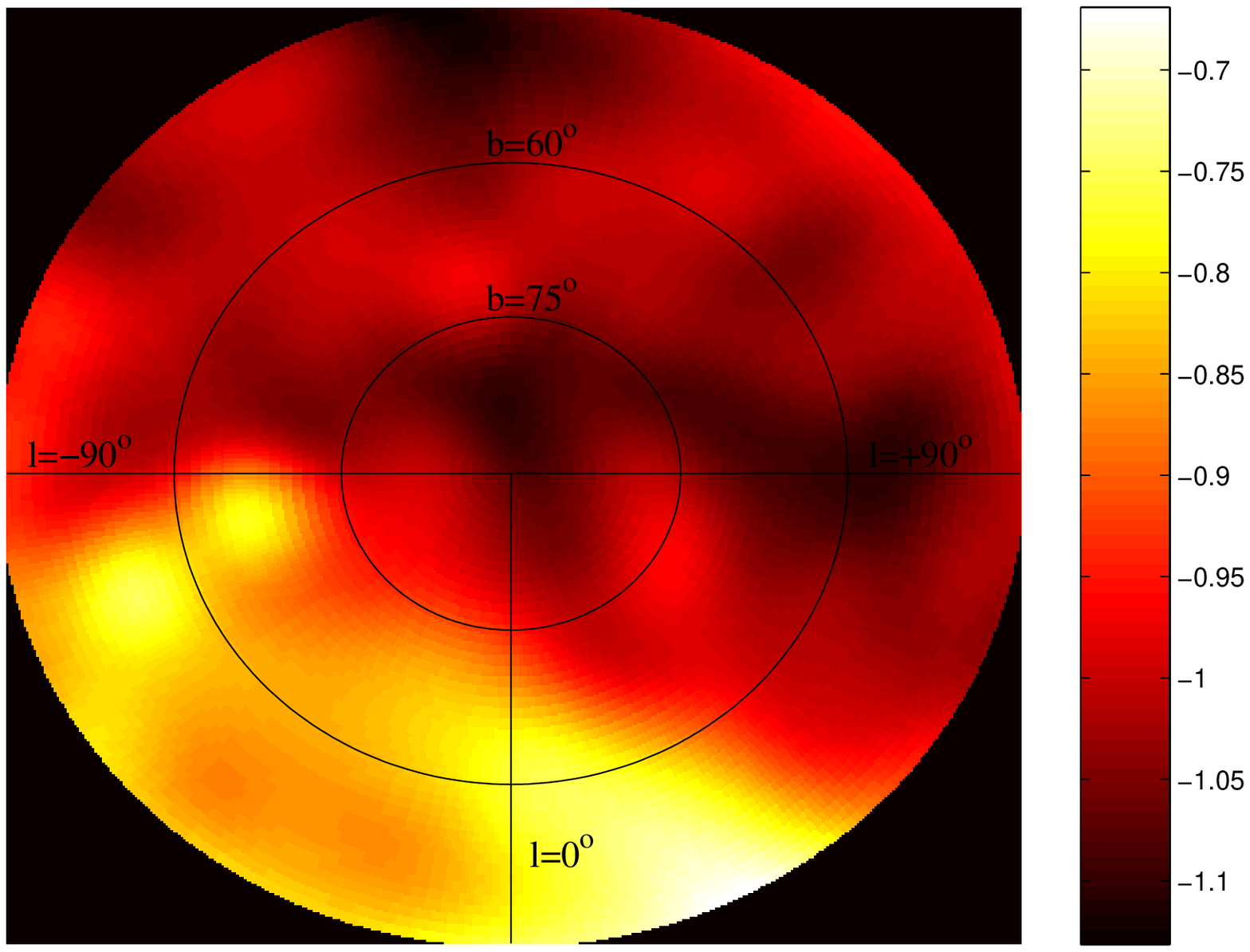}
\epsscale{5.06}
\plottwo{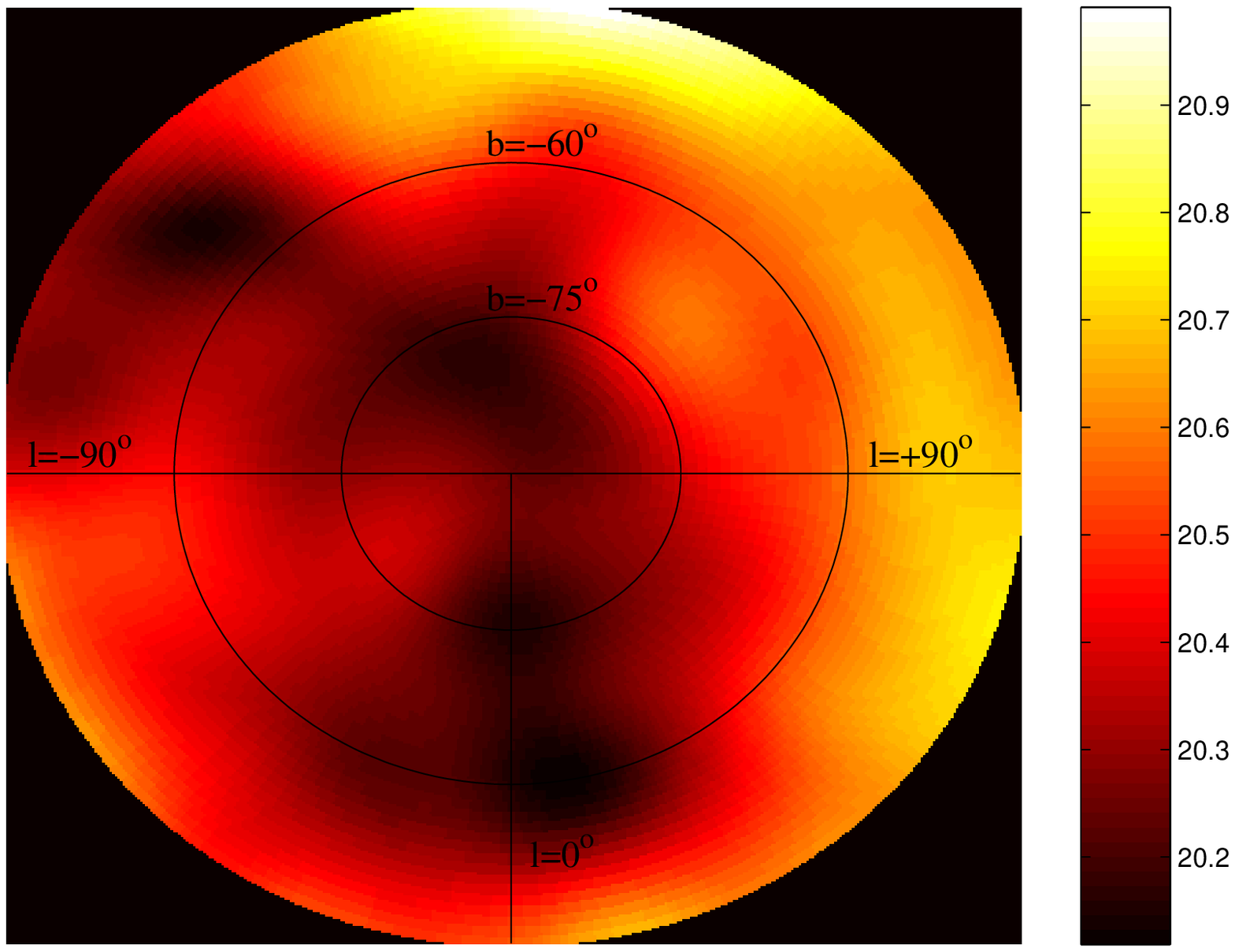}{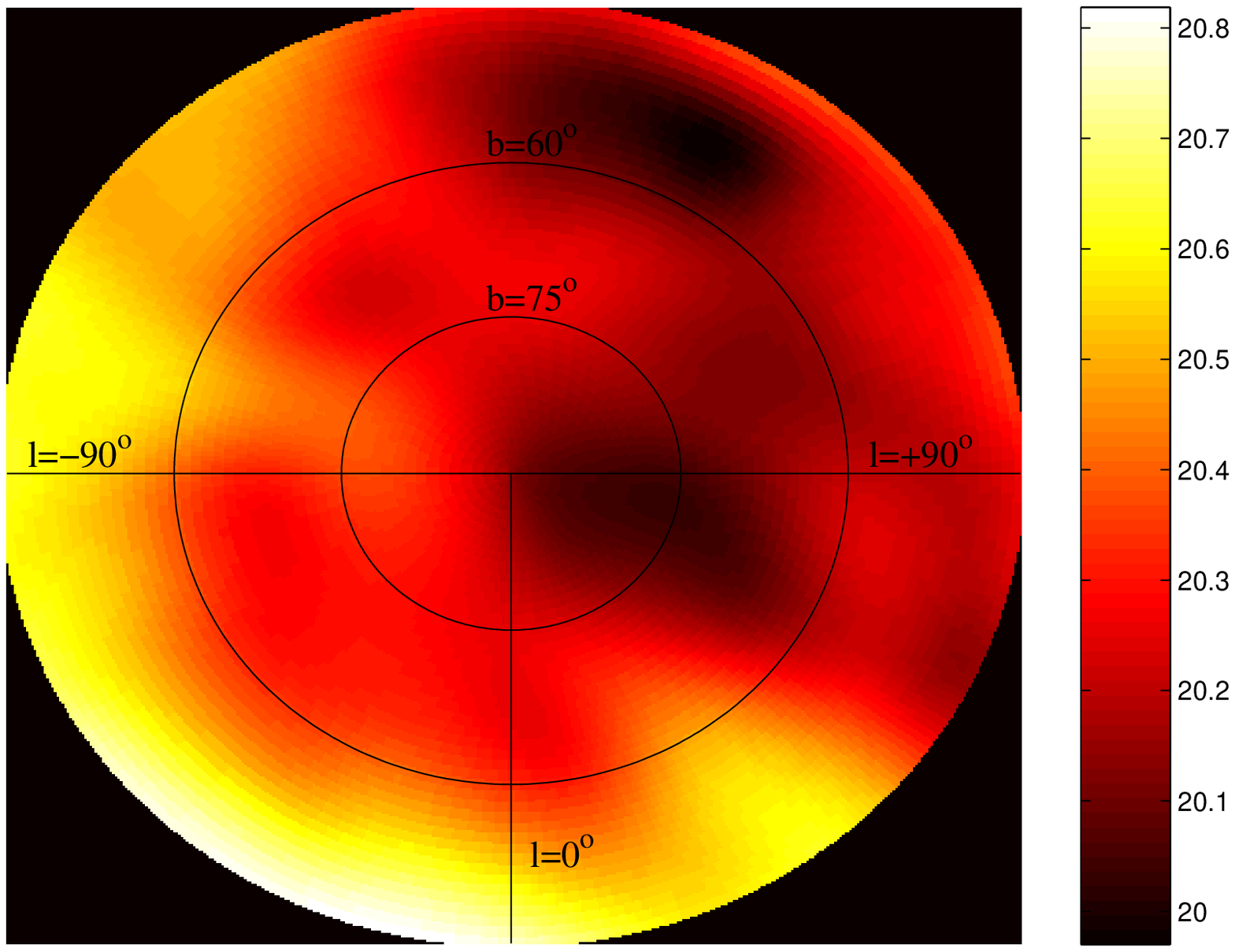}
\caption{ 
Images of the south (\emph{left panels}) and the north (\emph{right panels}) 
Galactic polar regions. 
Shown (in $\log_{10}$ scale) are the \gama-ray intensity (above $100\MeV$, 
in units of $\Iunits$, \emph{upper panels}), the synchrotron effective 
temperature at $23\GHz$ (in mK, \emph{middle panels}), and the \HI column 
density inferred from $21\cm$ line emission (in $\mbox{cm}^{-2}$, 
\emph{bottom panels}). 
The \gama-ray images include point sources such as 3EG J0210-5055 
($l=276.10\de$, $b=-61.89\de$). 
}
\label{fig:polar_images}
\end{figure}

Figure~\ref{fig:latitude_profiles} shows the average latitude profile of the 
\gama-ray data, at latitudes $|b|>42\de$, for both the north and south 
Galactic hemispheres. 
The data points constituting the latitude profiles were obtained by removing 
the point sources, smoothing with an EGRET PSF-like Gaussian filter 
(see \S \ref{subsec:gamma_ray_data}), 
and averaging over $4\de$ concentric rings about the Galactic poles. 
For comparison, the figure also shows the corresponding latitude profiles 
extracted from two selected Galactic tracers 
(see \S \ref{subsec:galactic_tracers_data}): 
a high-frequency synchrotron map (in $23\GHz$, tracer 2), and an \HI column 
density map based on $21\cm$ line-emission (tracer 3). 
For illustration, Figure~\ref{fig:polar_images} presents images of the south 
polar region (SPR, $b<-45\de$) and of the north polar region (NPR, $b>45\de$), 
for the data of EGRET and the two selected Galactic tracers. 

The \gama-ray latitude profile shown in Figure~\ref{fig:latitude_profiles} 
exhibits a characteristic $\sim 1/\sin|b|$ behavior, typical of an emitter 
with a disk-like geometry, at latitudes $42\de<|b|\mla 70\de$. 
At higher latitudes ($|b|\mga 70\de$), the profile steepens, decreasing 
approximately linearly with $|b|$ as the line of sight approaches the poles.  
This intensity decline corresponds to the dark regions around the poles, 
evident at the centers of the \gama-ray images presented in 
Figure~\ref{fig:polar_images}. 
The relative \gama-ray brightness of the northern hemisphere compared to the 
southern hemisphere persists at all latitudes $|b|>42\de$, although the 
difference between the two hemispheres, larger than $10\%$ near the poles, 
gradually diminishes at lower latitudes. 
Notice the general agreement between the \gama-ray profiles of the two 
Galactic hemispheres, lending credence to the data and to the significance of 
these profiles.  

Before discussing the data quantitatively, we point out two qualitative 
consequences of figures \ref{fig:latitude_profiles} and \ref{fig:polar_images}.
First, there are qualitative similarities between the high-latitude \gama-ray 
profile and the profiles of the two Galactic tracers. 
In particular, the $23\GHz$ synchrotron profile also features a relative 
brightening of the northern hemisphere compared to the south. 
Partial resemblance, but also several differences, can be identified between 
the three images of each polar region shown in Figure~\ref{fig:polar_images}, 
indicating strong correlations between the three data sets. 
Second, note that the \gama-ray profiles are highly inconsistent with 
isotropy, \emph{especially} near the poles, but roughly agree with the 
expected signature of Galactic emission at all latitudes. 
This result precludes the possibility that the Galactic contribution to the 
polar intensity is negligible with respect to an extragalactic isotropic 
component, and supports the conclusion that the polar Galactic component is at 
least comparable to the EGRB flux.

\subsection{Fitting the Average \gama-ray Latitude-Profile}
\label{subsec:high_latitude_fit}

We now explore the consequences of the aforementioned similarities between the 
latitude profile of the \gama-ray data and the profiles of the Galactic 
tracers. 
In Appendix \ref{sec:galactic_model} we argue, in accord with previous 
studies, that the Galactic \gama-ray emission arises mainly from three 
radiative processes involving CRs: inverse-Compton scattering of background 
photons by electrons, electron bremsstrahlung, and proton-nucleon scattering. 
The first component is expected to be highly correlated with synchrotron 
emission, whereas the latter two components should be better correlated with 
the column density of interstellar gas. 
Under certain conditions, e.g. small fluctuations in the distributions of CRs 
and gas on the relevant scales, global properties of the Galactic \gama-ray 
emission, such as its latitude profile, may be postulated to be roughly 
proportional to a \emph{weighted sum} of the properties of synchrotron 
emission and of the gas column density. 
The ratio between the weights of these two components, $(w_{syn}:w_{gas})$, 
is uncertain and may well vary with latitude. 
However, we demonstrate in Appendix \ref{sec:galactic_model} that at high 
latitudes this ratio should be roughly constant, and can be estimated with 
simple, albeit somewhat extreme assumptions, to lie between $6:4$ and $8:2$. 

Motivated by this model, we attempt to simultaneously fit the measured 
\gama-ray high-latitude profiles of both hemispheres ($24$ data points with 
the resolution selected) as a weighted sum of two Galactic tracer profiles --- 
a synchrotron tracer and a gas tracer --- where the ratio between the two 
weights serves as a free parameter. 
We first employ the two Galactic tracers displayed in 
Figure~\ref{fig:latitude_profiles}: high frequency ($23\GHz$) synchrotron 
emission, and \HI column density based on $21\cm$ line emission. 
We find that the \gama-ray profile is very well fit by a weighted sum 
of the latitude profiles of the two selected Galactic tracers, if 
$w_{syn}:w_{gas}\simeq 7:3$. 
The best fit, displayed in Figure~\ref{fig:latitude_profiles}, is obtained 
when the ratio $w_{syn}:w_{gas}$ is allowed to vary slightly between the 
two Galactic hemispheres (i.e. using $2$ free parameters), giving $8:2$ in the 
north and $7:3$ in the south.  
Such small differences between the two hemispheres are to be expected, because 
of strong local features such as the north polar spur (loop 1), dominating the 
NPR (see Figure~\ref{fig:polar_images}).

\begin{figure}
\epsscale{1.0}
\plotone{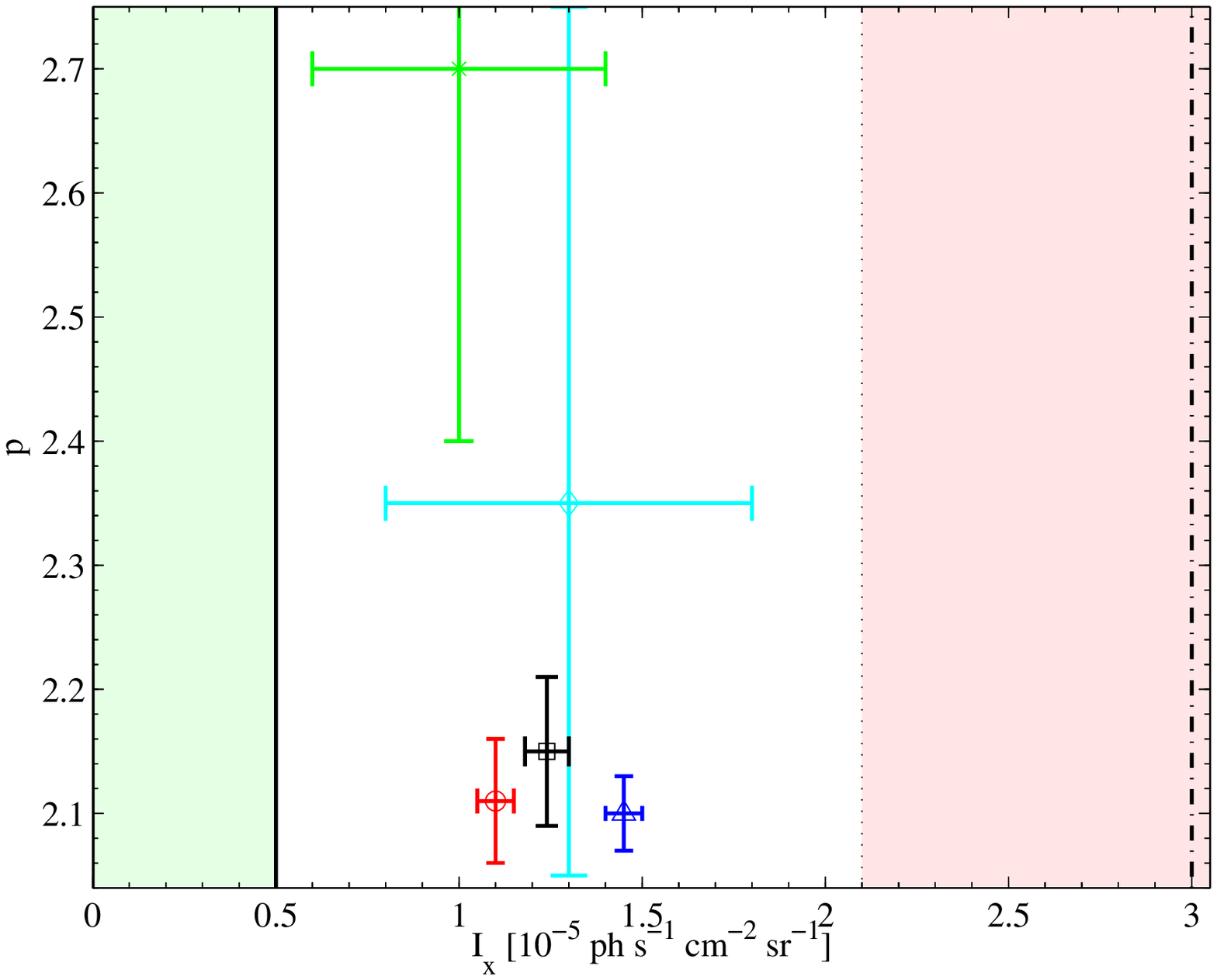}
\caption{ 
Various estimates of the EGRB flux $I_x$ and spectral index $p$. 
Shown are the $1\sigma$ error-bars estimated by 
Fichtel et al. (1978, based on correlations with $21\cm$ line emission, 
\emph{star}); 
Thompson \& Fichtel (1982, based on correlations with galaxy counts, 
\emph{diamond}); 
Osborne, Wolfendale \& Zhang (1994, based on correlations with $21\cm$ 
emission and the inner Galaxy-outer Galaxy asymmetry, \emph{circle}); 
Chen, Dwyer \& Kaaret (1996, based on correlations with $21\cm$ and $408\MHz$ 
emission, \emph{square}); 
and Sreekumar et al. (1998, based on a three-dimensional model, 
\emph{triangle}). 
The $1\sigma$ estimate of the EGRB flux obtained by Kraushaar et al. (1972, 
based on correlations with $21\cm$ emission) 
is shown as a shaded region around their best fit value 
(\emph{dash-dotted line}). 
The $3\sigma$ upper limit on the EGRB flux found in this work using the 
high-latitude ($|b|>42\de$) data profile is shown as a \emph{solid line}.  
}
\label{fig:EGRB_results}
\end{figure}

Next, the above two-component model was extended to include an additional, 
isotropic component, representing the extragalactic contribution to the 
\gama-ray latitude-profile. 
We find that the model fit to the data is not improved by adding such a 
component. In fact, a marginal improvement of the fit is achieved if such an 
isotropic component is not added to, but rather subtracted from the model.  
These findings thus lead to the conclusion that the latitude profile of the 
\gama-ray data shows no indication of an extragalactic, isotropic component, 
but can be well accounted for by a model-motivated combination of known 
Galactic components. 
Such a three-component fit, including an isotropic component, yields upper 
limits on the EGRB flux. 
The limits imposed by the north Galactic data, 
e.g. $I_x<0.2 \,(\atan86\CL)$ and $I_x<0.4 \,(\at98\CL)$, 
are slightly weaker than limits obtained from the southern hemisphere, 
and each hemisphere separately gives  
\beq  I_x<0.5 \,(\at99\CL) \fin \eeq
This result is shown in Figure \ref{fig:EGRB_results}, in comparison to previous estimates of the EGRB. 
In order to examine the sensitivity of this result to the Galactic tracers used, we repeat the analysis for various combinations of synchrotron and gas tracers. 
Generally, very similar results are obtained if the $21\cm$ based \HI column 
density map is replaced by other tracers of the gas, such as \Halpha emission 
(tracer 4) or the dust column density map (tracer 5, assuming a uniform 
gas-to-dust ratio), although the associated fit to the data is not as good. 
Replacing the high frequency ($23\GHz$) synchrotron data with the low 
frequency map ($408\MHz$, tracer 1) leads to a far worse fit, mainly because 
of local bright features present near the south Galactic pole in the 
$408\MHz$ map. 
However, the main result --- no evidence for an isotropic component --- 
remains unchanged. 

\subsection{Steepness of the Latitude Profile}
\label{subsec:gamma_ray_steepness}

One may adopt a more phenomenological approach to the analysis of the high 
latitude profile, free of even the basic Galactic model considerations 
presented above. 
The average \emph{smoothed} \gama-ray intensity found within $4\de$ of the 
Galactic polar caps, $I_s(|b|> 86\de)\simeq 1.48\pm 0.09$, is very close to 
recent estimates of the EGRB flux, e.g. $I_x=1.45\pm0.05$ \cite{Sreekumar98}.
For the present analysis, let us ignore the fact that smoothing the data has 
artificially increased the polar intensity (see \S \ref{sec:galactic_poles}), 
because the Galactic tracers were identically smoothed and because we are 
interested here only in the global properties of the data. 
If we adopt previous EGRB flux estimates, such as presented above, as a null 
hypothesis, we are lead by the low polar intensity to deduce that the Galactic 
foreground towards the poles is negligible, constituting no more than 
$\sim10\%$ of the EGRB flux. 
However, the \gama-ray intensity measured at slightly lower latitudes is much 
higher, because the latitude profile is steep. 
For example, the intensity at $|b|\sim 60\de$ is roughly $40\%$ higher than 
the polar intensity, at virtually all longitudes. 
Hence, this null hypothesis implies a very steep Galactic \gama-ray 
foreground, with an intensity that varies by a factor of $\sim 4$ between 
$|b|\sim90\de$ and $|b|\sim60\de$. 
Such a Galactic profile is inconsistent with known components of the Galaxy 
and with measured Galactic tracers, thus ruling out the above null hypothesis. Evidently, this is a restatement of the conclusion, derived above and used in 
\S \ref{sec:galactic_poles}, that the Galactic contribution to the polar 
intensity is not negligible. 

We may now constrain the EGRB flux differently, by finding out the maximal 
isotropic flux that when added to reasonable Galactic profiles, 
may produce the steep \gama-ray profile observed. 
Quantitatively, we model the measured \gama-ray latitude profile as a linear 
combination of an isotropic component and a single Galactic tracer, with the 
ratio between the weights of the two components serving as a free parameter. 
We are interested here in the global properties of the latitude profiles, 
i.e. their steepness, and thus focus on very high latitudes, $|b|>58\de$, 
where the effect of local Galactic features is minimal. 
Nonetheless, various features present in the data yield results that are 
sensitive not only to the tracer used, but also to the Galactic hemisphere 
studied. 
The \emph{highest} isotropic flux thus found is obtained with the \HI column 
density based on the $21\cm$ emission (tracer 3) applied to the southern 
Galactic hemisphere, giving $I_x = 0.50\pm 0.13$. 
However, all other Galactic tracers yield lower estimates for the isotropic 
component, in most cases ruling out a significant EGRB flux. 
For example, the \emph{best} fit to the data is obtained with the high 
frequency synchrotron profile (tracer 2) applied to the northern hemisphere, 
giving $I_x = 0.0\pm0.2$.
Hence, this method yields no conclusive estimate of the EGRB flux, but one can 
still use the highest isotropic component estimated above to impose upper 
limits on the EGRB flux, 
\beq I_x<0.6 \,(\at78\CL) \AND I_x<0.8\,(\at99\CL) \coma \eeq 
with evidence for a flux far lower than this. 

To conclude this section, the high-latitude profile of the \gama-ray data, 
a simple but illuminating measure of the \gama-ray sky, is found to exhibit 
strong \emph{Galactic} features, indicating that the Galactic foreground is 
strong at all latitudes, even near the Galactic poles. 
The \gama-ray profile is well fit by a model-motivated combination of a 
Galactic synchrotron tracer and a Galactic gas column density tracer, 
imposing an upper limit $I_x<0.5\,(\at99\CL)$. 
The mere steepness of the latitude profile independently gives, when compared 
with known Galactic tracers, $I_x<0.8\,(\at99\CL)$.

\section{All-Sky Analysis}
\label{sec:all_sky_analysis}

After studying parts of the \gama-ray sky where the Galactic foreground is 
relatively low, we now turn to an all-sky analysis of the data. 
An all-sky separation of the EGRB from the strong Galactic foreground requires a model for Galactic \gama-ray emission, which in turn depends on the 
uncertain distributions of CRs, gas and radiation fields through the Galaxy. 
One may either use three-dimensional models of these components, constructed 
from various observations of their radiative consequences, or the observed 
two-dimensional sky maps directly. 
Whereas three dimensional models are more physically motivated and can predict the consequences of interactions between the different Galactic components, 
they are often oversimplified, suffering from several uncertainties, and often assuming some symmetry relations in translating two-dimensional sky maps to a 
three-dimensional Galactic model. 
An example of an EGRB study using such a three-dimensional model is reviewed 
in \S \ref{sec:Sreekumar98}. 
In this section we focus on the use of two-dimensional sky maps of various 
Galactic tracers in order to measure the EGRB. 
The basic motivation is to identify and subtract from the measured \gama-ray 
map, a component which is highly correlated with a Galactic tracer, thus 
eliminating some of the Galactic foreground and imposing an upper limit on 
the radiation which is not associated with the Galaxy. 
Two-dimensional Galactic tracers provide little information regarding the 
outcome of interactions between different Galactic components because of 
projection effects, 
thus imposing only loose, uncontrollable upper limits on the EGRB flux. 
On the other hand, this method is free from a-priori assumptions regarding 
the \gama-ray emission processes and the distributions of the Galactic 
components involved and can take advantage of the high level of detail present in the tracers. 
Such methods have been extensively used in previous studies of the EGRB. 

We begin in \S \ref{subsec:previous_correlation_studies} with a brief overview of previous studies which have employed correlations between \gama-ray maps and various Galactic tracers. 
In \S \ref{subsec:method} we describe the method used in our all-sky analysis, emphasizing its limitations. 
The results are presented and discussed in \S \ref{subsec:all_sky_results}. 
In particular, we examine the sensitivity of the results to the Galactic 
tracers used, and to the part of the sky being examined.

\subsection{Previous Correlation-Based Studies} 
\label{subsec:previous_correlation_studies}

Most past searches for a diffuse EGRB have correlated the \gama-ray data with 
various Galactic tracers in order to subtract the Galactic foreground and 
identify the extragalactic component. 
The results of such studies are summarized in Figure \ref{fig:EGRB_results}. 
Two important points should be kept in mind. 
First, with limited knowledge of the three-dimensional distributions of the 
Galactic components and the correlations between them, the results of 
correlation-based methods should be considered only as upper limits to the 
true EGRB flux. 
Second, the results of such methods are not stable but rather depend on the 
choice of Galactic tracer or tracers used and on the part of the sky being 
examined. 
As we show in \S \ref{subsec:all_sky_results}, this dependence is strong, 
rendering the results of such methods inconclusive. 
For illustration, we present two examples of correlation-based methods previously applied to the EGRET data. 

Osborne, Wolfendale \& Zhang (1994) have analyzed the early EGRET data 
(phase 1) in search of the EGRB. 
First, they correlated the \gama-ray data with the \HI column density found 
from a $21\cm$ survey \cite{Stark92} and removed the correlated component. 
Next, they used the difference between the remaining \gama-ray intensity of 
the inner Galaxy and the outer Galaxy in order to identify and remove an 
additional Galactic contribution, interpreted as inverse-Compton emission, 
according to the model of Chi et al. (1989).  
The remaining, roughly isotropic component was identified as the EGRB, 
with flux $I_x=1.10\pm0.05$ and spectral index $p=2.11\pm0.05$. 
Osborne et al. report a large systematic difference between their EGRB flux 
estimates based on the northern Galactic hemisphere and on the southern 
hemisphere, and suggest that the difference is of an extragalactic nature.
 
Chen, Dwyer \& Kaaret (1996) have fit the data of EGRET (phases 1 and 2) with 
a \emph{three}-component model, consisting of a component which is linear in 
the \HI column density inferred from $21\cm$ line emission \cite{Dickey90}, 
a component linear in synchrotron emission from a $408\MHz$ survey 
\cite{Haslam82}, and an isotropic extragalactic component. 
They found a higher, slightly softer EGRB than found by Osborne et al.: 
$I_x=1.24\pm0.06$ and $p=2.15\pm0.06$.  
In \S \ref{subsec:method} we discuss the reliability of such multi-component 
models. 

Note that these correlation-based studies of the EGRET data disagree with the 
most recent study of the EGRB by Sreekumar et al. (1998), discussed in 
\S \ref{sec:Sreekumar98}, where an elaborate three-dimensional model of 
Galactic \gama-ray emission \cite{Bertsch93,Hunter97} was applied to the EGRET 
data. 
The EGRB flux found by Osborne et al. (1994) and by Chen et al. (1996) are 
lower than found by Sreekumar et al. (1998) at a $7\sigma$ and at a 
$3.5\sigma$ level, respectively.

\subsection{Method}
\label{subsec:method}

The separation of an extragalactic \gama-ray component from the strong 
Galactic foreground relies crucially on the assumption that the EGRB is 
isotropic when integrated over large enough scales, say $\ga5\de$. 
The \gama-ray data is thus $\chi^2$ fitted to a model in which the observed 
\gama-ray background is the weighted sum of an isotropic component and a 
Galactic tracer, with the ratio between the two weights considered as a free 
parameter. 
The best fit value for the isotropic component may then be used as an 
approximate upper limit to the EGRB flux, provided that the resulting fit is 
good, the two weights are well determined and their values are physically 
sensible. 
We stress that merely a loose upper limit can be obtained in this way, because the Galactic \gama-ray emission is unlikely to be proportional to any Galactic tracer or combination of tracers, a problem further complicated by projection 
effects. 
Only a good understanding of the three-dimensional distributions of the 
Galactic components would enable one to measure the tightness of this upper 
limit or to improve it. 
As discussed in Appendix \ref{sec:galactic_model}, the present understanding 
of the distributions of these Galactic components is insufficient for this 
purpose. 
Finally, this method entails the danger of underestimating the extragalactic isotropic flux in situations where the tracer used contains an artificial isotropic component, e.g. because of insufficient elimination of an extragalactic contribution to a synchrotron radio map or an over-smoothed 
tracer. 
Therefore, special care has been taken to remove any spurious isotropic 
artifact from the Galactic tracers analyzed. 

In principle, the \gama-ray sky map obtained by subtracting a component 
linear in a Galactic tracer may be used again, repeating the process. 
The modified map may be correlated with another Galactic tracer, and an 
additional component of the \gama-ray sky, allegedly associated with different 
components of the Galaxy, may be subtracted, thus imposing tighter limits on 
the EGRB flux. 
However, this method would be effective only if the Galactic \gama-ray 
emission is linear in the Galactic tracers used on the relevant scales. 
A sequential subtraction of tracer-correlated components further requires the 
tracers to be uncorrelated, and the errors introduced by each two-component 
model fit are cumulative. 
However, since this method has been previously used \cite{Chen96}, we examine 
it quantitatively here. 

In the first stage of our analysis, the data is cleaned from various sources 
of noise, such as \gama-ray point sources or extragalactic background in the
synchrotron tracers (see \S \ref{subsec:galactic_tracers_data}). 
Next, the sky is divided into $N$ regions of approximately constant solid 
angle, the data is accordingly binned, and the linear correlations $r_k$ 
between the \gama-ray data and each tracer $k$ are calculated. 
A simple Monte-Carlo simulation, based on reshuffling the bins in one of the 
data sets, to estimate the probability of randomly finding a linear 
correlation coefficient larger than $r_k$, usually yields probabilities 
similar and often slightly lower than one would predict for $r_k$ by assuming 
that the underlying distributions of the two data sets jointly form a 
two-dimensional Gaussian (bi-normal) distribution. 
In order to confidently identify a cross correlation signal between two data 
sets with no a-priori assumptions regarding their underlying distributions, 
we turn to more robust estimators of the correlation. 
We thus evaluate the Spearman (non-parametric) rank-order correlation 
coefficient $s_k$, found by replacing each data element by its rank within its 
data set and calculating the linear correlation between the two sets of ranks. 

We exclude the Galactic plane ($|b|<6\de$) and the inner Galaxy ($|b|<42\de$, 
$|l|<60\de$) from our analysis, since these regions are strongly dominated by 
Galactic emission and because the Galactic composition of these regions is 
probably not captured by the Galactic tracers we use (e.g. we do not employ 
a tracer of the \HH gas, which is highly concentrated at low latitudes).  
We divide the remaining sky into bins of approximately constant solid angle, 
with characteristic angular scale $\theta=12\de$, roughly corresponding to the 
$67\%$ containment of the EGRET PSF, and repeat our analysis for 
$\theta=24\de$ and $\theta=6\de$. 
The maps of the \gama-ray sky and of the Galactic tracers are binned 
accordingly, after being cleaned and smoothed with a filter function designed 
to imitate the EGRET point spread function 
(see \S~\ref{subsec:gamma_ray_data}). 
We assume a fractional systematic error $\Delta$ in each \gama-ray bin, in 
addition to the statistical errors, and repeat our analysis for various 
values of $\Delta$ ($5\%$, $10\%$ and $20\%$). 

Although a multi-frequency analysis of the \gama-ray data can, in principle, 
provide a better estimate of the extragalactic component, we do not adopt such 
an approach in this study, because of three reasons: 
(i) the EGRB is found to be much weaker than the Galactic foreground and modelling the spectral data can only lower our upper limit to the EGRB flux, 
(ii) a-priori information regarding the EGRB spectrum is poor, 
and (iii) there are poorly understood spectral variations in the Galactic 
foreground across the sky, even in regions of evident Galactic dominance 
\cite{Hunter97}. 
Furthermore, previous studies of the EGRET data have reported a hard EGRB spectrum with spectral indices $p\simeq 2.0-2.2$, similar to the expected spectrum of both the bremsstrahlung and the inverse-Compton components of the Galactic \gama-ray emission in energies around $100\MeV$
[see \S \ref{subsec:previous_correlation_studies}, Hunter et al. (1997), 
Sreekumar et al. (1998) and Appendix \ref{sec:galactic_model}]. 
For such an EGRB spectrum, the spectral information can not be used to 
distinguish between the EGRB and most of the Galactic foreground, further 
justifying our emphasis on the spatial dependence of the data.

\subsection{Results}
\label{subsec:all_sky_results}

Most of the two-component models, involving an isotropic component and a 
component linear in a Galactic tracer, admitted poor fits. 
Acceptable $\chi^2/\nu$ values, where $\nu=N-1$, were obtained only for 
$\Delta\ga 10\%$. 
The best fit is obtained using the $21\cm$ tracer, giving an isotropic flux 
$I_{iso}\simeq 1.51\pm0.04$ 
(with $\chi^2/\nu=0.76$ for $\theta=12\de$ and $\Delta=10\%$). 
Similar procedures were indeed used by previous authors 
\cite{Kraushaar72,Fichtel78,Thompson82} to obtain comparable results. 
For example, Fichtel et al. (1978) have analyzed the SAS-2 data assuming that 
the Galactic foreground is proportional to the \HI column density deduced from 
$21\cm$ line emission, finding $I_x\simeq 1.0\pm0.4$ and 
$p\simeq2.7_{-0.3}^{+0.4}$. 
Needless to say, such estimates should be considered only as an upper limit 
to the true EGRB flux. 
Moreover, some two-component models involving other Galactic tracers, provide 
lower estimates of the isotropic flux, with a similar significance. 
For example, the $23\GHz$ tracer gives a marginally acceptable fit for an 
isotropic flux $I_{iso}\simeq0.77\pm0.06$ (with $\chi^2/\nu=1.81$), 
and the synthetic large-scale map (tracer 7) produces an acceptable fit with 
a smaller isotropic component, $I_{iso}\simeq0.62\pm0.06$ 
(with $\chi^2/\nu=1.26$).
Evidently, two-component models involving different tracers lead to 
contradicting results. 

Applying the method to different, limited, parts of the sky indicates that the 
situation is more complicated. 
Correlations of the \gama-ray data with the tracers vary across the sky, 
producing different estimates for $I_{iso}$ in different regions. 
The results exhibit both a latitude dependence, and a $1.5\sigma-4\sigma$ 
difference between the two Galactic hemispheres 
(see also Osborne et al. 1994; Chen et al. 1996).  
Thus, Galactic \gama-ray emission is not simply proportional to any single 
tracer, implying that the method is sensitive to variations in the Galactic 
composition across the sky and to strong local features 
(e.g. the north polar spur). 
As an example, the correlations between the \gama-ray data and the $21\cm$ map at high ($|b|>42\de$) latitudes are stronger in the northern hemisphere, 
giving $I_{iso}\simeq1.35\pm0.10$, lower by $\sim1.5\sigma$ than in the south. A better fit is obtained with the synthetic large-scale map, favoring no isotropic component at all at high latitudes, in either hemisphere. 
Hence, if the Galactic \gama-ray emissivity is assumed proportional to the 
large-scale structure of the Galaxy, the disk-like profile of the 
high-latitude data suggests a negligible isotropic component. 
Considering the spatial dependence of the results and their sensitivity to the 
tracer used, it is not surprising that early studies, where the Galactic 
\gama-ray foreground was implicitly assumed linear in a single tracer, 
have not converged on an estimate of the EGRB flux. 

A higher stability across the sky is obtained when the \gama-ray data is 
modeled as a linear combination of several Galactic tracers, although this 
approach is less rigorously justified. 
As an example, we model the \gama-ray intensity as a linear combination of 
the non-synthetic tracers (1)-(5). 
The basic features found in the two-component models are still present: 
a latitude dependence and a $\sim 1\sigma$ difference between the isotropic 
flux deduced from each hemisphere, 
but the differences are now smaller because the relative tracer weights are 
allowed to vary dramatically between different parts of the sky. 
For example, the best fit attributes $\sim20\%$ of the high latitude \gama-ray 
intensity to a component linear in the $23\GHz$ map if the two hemispheres are 
examined simultaneously, but attributes nothing to this component when each 
hemisphere is considered separately, demonstrating the inherent weakness of 
the method.  
The best fit is obtained from the northern hemisphere at high latitudes, 
giving $I_{iso}\simeq1.0\pm0.1$ as an upper limit to the EGRB flux. 

The isotropic flux found by the multi-component model, lower than obtained 
above by some two-component models, suggests that the latter remove only 
part of the foreground.  
It is important to note, however, that we have no indication that the true 
value of the EGRB flux is close to estimates of $I_{iso}$ found in this 
section. 
In \S \ref{subsec:method} we have outlined the subtleties involved in the 
methods employed, in particular the uncontrollable nature of the assumptions: 
there is no guarantee that the above estimates do not merely reflect 
projection effects, systematic errors lurking in the data, and complicated 
(non-linear) cross-correlations between the Galactic tracers. 
In order to demonstrate this point, we repeat the multi-component fit, 
replacing the dust map (tracer 5) with the synthetic map featuring the 
large-scale structure of the Galaxy. 
The results favor no isotropic component at all at high latitudes, with a 
better fit to the data than obtained previously. 
A large fraction of the high-latitude \gama-ray map is found to be roughly 
proportional to the large-scale structure of the Galaxy. 

In conclusion, we have followed previous studies and used correlations with 
Galactic tracers as a tool for removing the Galactic foreground and uncovering 
the EGRB. 
We have found such correlation methods to be potentially misleading: 
the results vary significantly as a function of the tracers used and the part 
of the sky under consideration. 
Our analysis recovered the results of previous studies and suggested an upper 
limit $I_x\la 1.0$; however it also indicated higher or lower values for some 
models. 
For example, if the \gama-ray emissivity is assumed to follow the large-scale 
structure of the Galaxy, an EGRB flux $I_x\la0.6$ is implied, with indications 
for a much lower flux. 
These results lead to two important conclusions. 
First, the variation of the results across the sky indicates that the Galactic 
foreground is not linear in any combination of tracers, but reflects the 
interactions between various components and local features. 
Second, the EGRB flux seems \emph{low} with respect to the Galactic foreground 
at all latitudes. 
Hence, the \gama-ray data can be entirely ascribed to Galactic emission, with 
reasonable assumptions regarding the distributions of the Galactic components, 
precluding an unequivocal measurement of the diffuse, isotropic EGRB.

\section{Previous Model-Based Study of the EGRET Data}
\label{sec:Sreekumar98}

\subsection{Description}

Finally, we turn to discuss the most recent and comprehensive analysis of the 
EGRET data (cycles 1-3), performed by Sreekumar et al. (1998) using an 
elaborate model for the Galactic \gama-ray foreground 
\cite{Bertsch93,Hunter97}. 
This model, like most of its predecessors, attributed the Galactic high-energy 
\gama-ray emission to interactions of CR electrons and protons with the gas 
and the radiation fields within the Galaxy, 
with little contribution from point sources. 
The distributions of the Galactic components were inferred from various direct 
and indirect methods. 
Thus, the distribution of \HI was deduced from observations of $21\cm$ line 
emission \cite{Dickey90}, 
the \HH distribution was based on \CO emission at $115\GHz$ \cite{Dame87}, 
and the \HII distribution was inferred from the pulsar-based model of 
Taylor \& Cordes (1993). 
However, The distributions of the remaining components --- CR electrons 
and protons and low energy radiation fields (mainly optical and infrared) --- 
were not well determined, introducing significant foreground uncertainties, 
especially at high latitudes (see Sreekumar et al. 1998 and 
Appendix \ref{sec:galactic_model}). 

The CR electron spectrum was thus calculated by propagating a simple power-law 
spectrum (with index $p=2.42$) through a leaky-box model \cite{Skibo93}, 
adjusted to agree with direct measurements carried out on top of the Earth 
atmosphere, and with radio observations towards the Galactic poles, 
assuming a strong Galactic magnetic field $B=10.5\muG$ 
(see Figure \ref{fig:CR_distribution}). 
The CR proton spectrum was based on direct measurements. 
The distributions of both CR electrons and protons through the Galaxy were 
assumed correlated with the distribution of gas, 
and modeled by convolving the total gas distribution (\HImin, \HH and \HIImin) 
with a two-dimensional Gaussian filter of HWHM $r_0=1.76\kpc$ 
(Hunter et al. 1997 and the references therein). 
The distributions and the energy densities of the low energy radiation fields, 
necessary for a calculation of the inverse-Compton component of the Galactic 
\gama-ray emission, were based on the calculations of 
Chi \& Wolfendale (1991). 

Sreekumar et al. have fit the EGRET data as the sum of 
a Galactic \gama-ray foreground, calculated according to the above model and 
normalized by a factor $B\simeq1$, and an isotropic, extragalactic component. 
The resulting isotropic intensity was estimated as $I_{iso}=1.45\pm0.05$, 
although the factor $B$ varied by $5\%-10\%$ between different energy bands 
below $100\MeV$ and by $>30\%$ for energy bands above $1\GeV$.
The extragalactic component was then analyzed. 
The Galactic plane and the inner Galaxy were excluded from this analysis, 
because of extreme domination of the Galactic emission and difficulties in 
modeling this emission. 
The remaining sky was divided into $36$ bins of angular size 
$\sim20\de-30\de$. 
The intensity of these bins was shown to agree with a Gaussian distribution, 
of mean intensity $\langle I\rangle \simeq 1.47$ 
and a $1\sigma$ value of $\Delta I = 0.33$.  
In this analysis, all bins were given equal weight, although the solid angle 
subtended by each bin varied by a factor of $\sim 2$, from $0.199$ radians 
at low latitudes to $0.421$ radians at high latitudes, thus overemphasizing 
the low-latitude bins.

\subsection{Analysis}

The EGRB flux found by Sreekumar et al., $I_x=1.45\pm0.05$, is higher than 
reported by previous studies of the EGRET data, carried out by Osborne et al. 
(1994) and by Chen et al. (1996), at a $7\sigma$ and at a $3.5\sigma$ level, 
respectively (see Figure \ref{fig:EGRB_results}). 
In \S \ref{sec:all_sky_analysis} we have demonstrated that correlation-based 
methods in general tend to favor lower values of $I_x$. 

As discussed in Appendix \ref{sec:galactic_model}, present models of Galactic 
\gama-ray emission are accurate only to within a factor of \emph{a few}, 
because of various uncertainties involving, 
(i) direct measurements of the local CR electron flux, 
(ii) settlement of such measurements with radio observations towards the 
poles, 
(iii) the distributions of the radiation fields and of \HIImin, 
and (iv) the spatial correlations between CRs, radiation fields and gas. 
These uncertainties have two important consequences. 
First, they introduce uncertainties in the variation of Galactic \gama-ray 
emission through the Galaxy and in the relative contributions of the processes responsible for it. 
The fitting procedure performed by Sreekumar et al. does not solve this 
problem, because one normalization factor ($B$) for the Galactic model can 
only fix the amplitude but not affect the flux variations across the sky and 
the ratio between the different Galactic emission components. 
Second, although the Galactic \gama-ray model can be adjusted to agree well 
with the low-latitude data \cite{Hunter97,Sreekumar98}, which is dominated by 
strong Galactic emission and was modelled with a relatively small inverse-Compton component, it clearly does not have the accuracy needed to identify the weak EGRB and separate it from the Galactic foreground. 
The ``difficulties in accounting for all the Galactic diffuse emission'' 
at low latitudes and towards the inner Galaxy are a manifestation of this 
conclusion: the Galactic model does not have sufficient accuracy to 
simultaneously account for the strong Galactic foreground at low latitudes and to enable elimination of the weaker foreground at high latitudes. 

The intensity fluctuations among the \gama-ray bins reported by Sreekumar 
et al. are \emph{stronger} than the estimated systematic and statistical 
errors in the intensity of each bin. 
For example, the intensity of the bin defined by $145\de<l<180\de$ and  
$-30\de<b<-10\de$ is only $0.89\pm0.17$, 
i.e. $3.3\sigma$ away from the estimated flux. 
Sreekumar et al. suggest that the intensity fluctuations between bins provide 
a measure of the anisotropy of the extragalactic radiation. 
However, the isotropy of the EGRB was implicitly assumed when fitting the data 
in search for the EGRB, and fluctuations of $>20\%$ on $>20\de$ scales are 
hard to produce in a physical model for the diffuse extragalactic background 
(e.g. Waxman \& Loeb 2000). 
If the EGRB is assumed isotropic on such scales, these fluctuations reflect 
systematic errors in the assumed model, much higher than reported. 
If these fluctuations are found to be correlated with Galactic tracers, 
some fraction of the reported EGRB can be identified as originating in fact 
from within the Galaxy, leaving behind a smaller component of a suspected 
extragalactic origin. 

\begin{figure}
\epsscale{1.0}
\plottwo{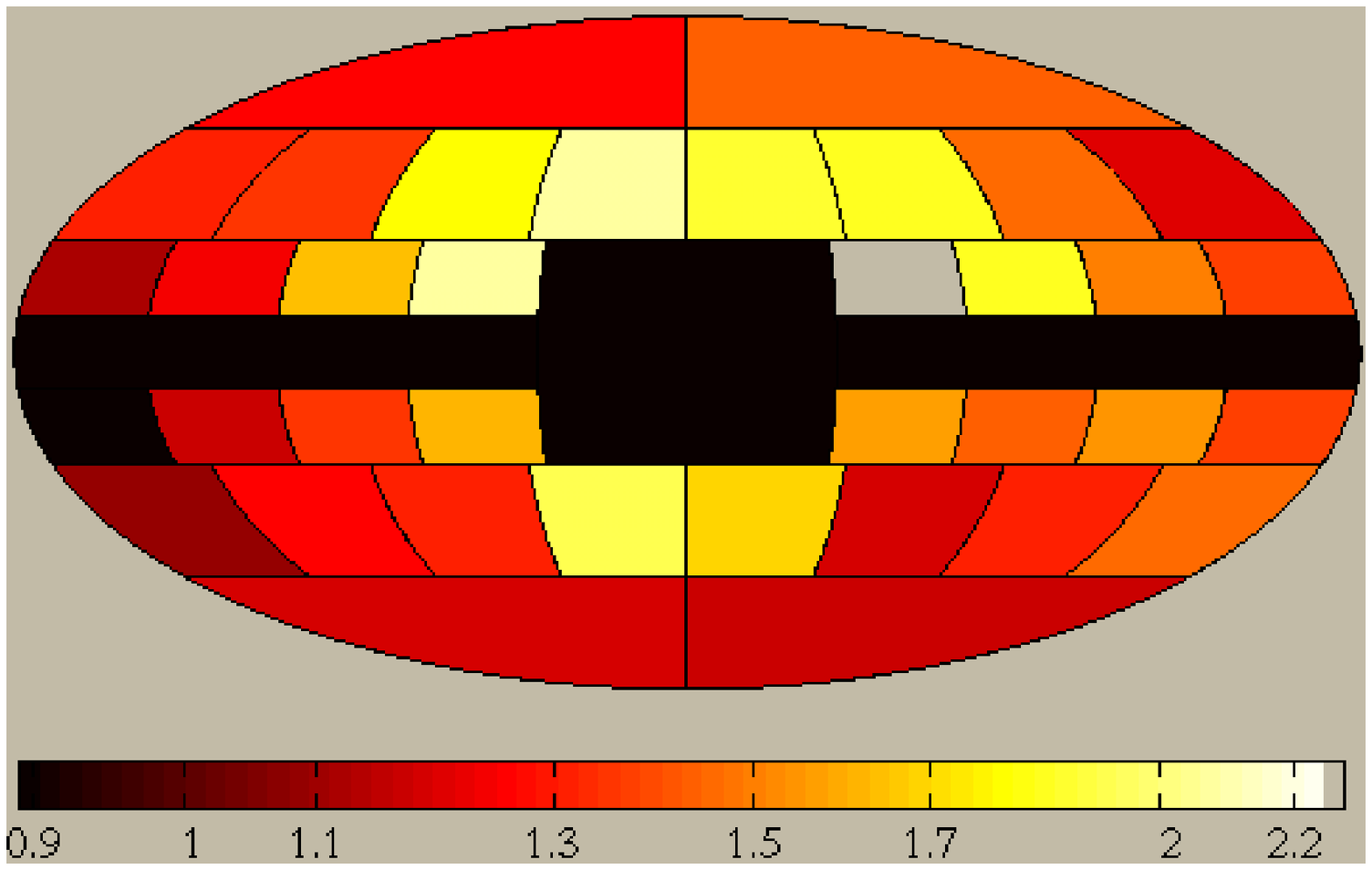}{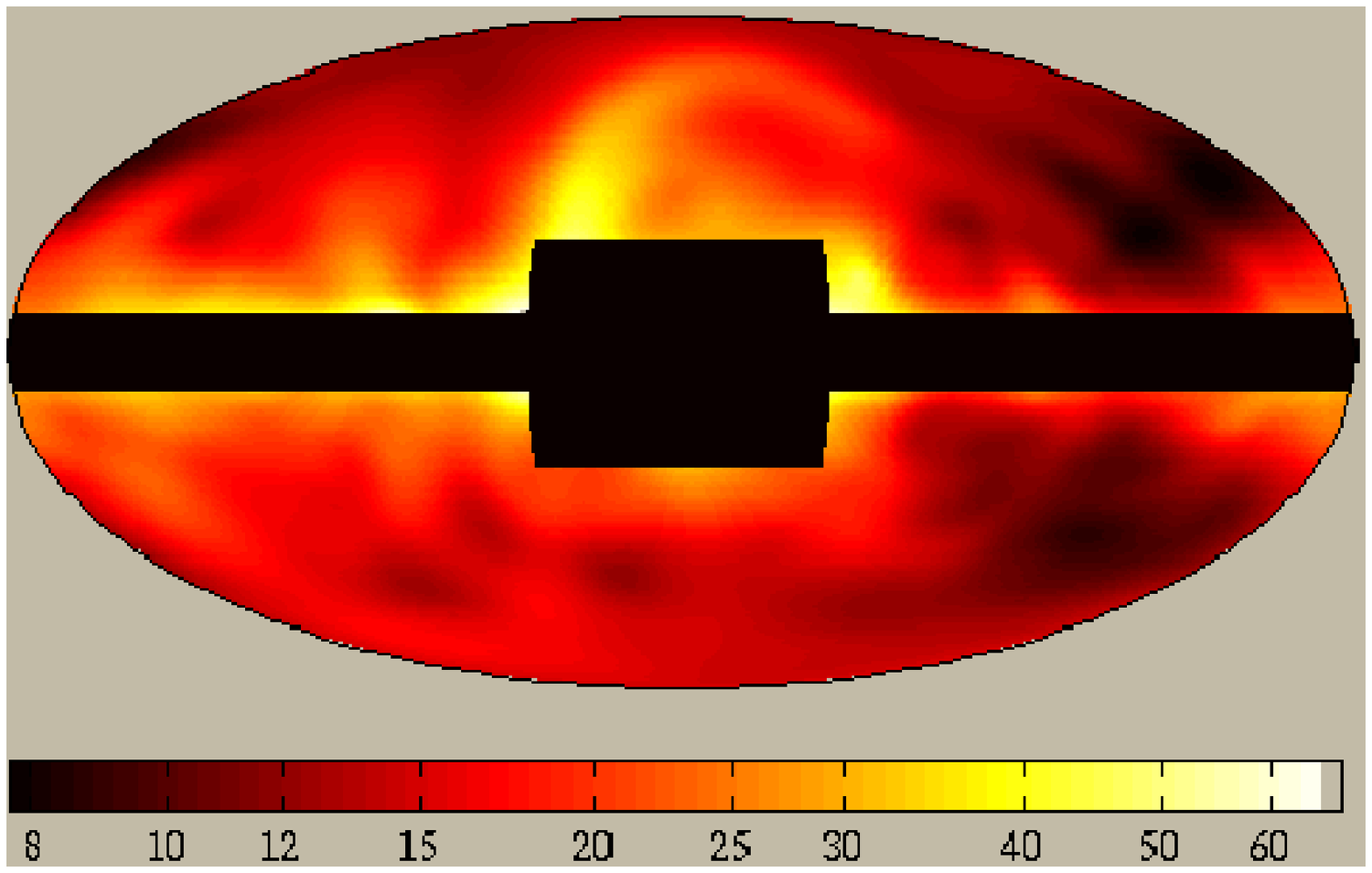}
\caption{ 
\newline
\emph{Left:} Extragalactic \gama-ray background intensity above $100\MeV$ 
found by Sreekumar et al. (1998), in a Hammer-Aitoff projection. 
The Galactic plane and the inner Galaxy were excluded because of difficulties 
accounting for the Galactic emission. Horizontal lines are plotted at latitudes $|b|=10\de,30\de$, and $60\de$. 
Color scale: $I\,[10^{-5}\Iunits]$.
\newline
\emph{Right:} All-sky survey in $408\MHz$ \cite{Haslam82}, cleaned from 
point sources and smoothed, in a Hammer-Aitoff projection.  
Color scale: $T\,[K]$.
}
\label{fig:maps_binned}
\end{figure}

We calculate the correlations between the EGRB map reported by Sreekumar et 
al. and the various Galactic tracers described in 
\S \ref{subsec:galactic_tracers_data}.  
Although the linear correlations of the EGRB map with the gas tracers are 
small or slightly negative, its correlations with the $408\MHz$ synchrotron 
map ($r_1=0.48$) and with the synthetic map of the Galactic large-scale 
structure ($r_7=0.28$) are significant.   
Figure \ref{fig:maps_binned} illustrates the situation, displaying the 
EGRB map of Sreekumar et al. alongside the $408\MHz$ map. 
Note that the low-latitude bins of Sreekumar et al. are smaller in terms of 
solid angle than the higher latitude bins: $\sim 75\%$ of the map (and $50\%$ 
of the sky) are found in medium and high latitude bins with $|b|>30\de$. 
By excluding the small part ($\sim25\%$) of the map represented by these 
low-latitude bins, we find that the EGRB map is correlated with \emph{all} 
the Galactic tracers, where the highest correlations are still found with the 
synchrotron and with the synthetic large-scale structure maps. 
One can explore the EGRB map of Sreekumar et al. with the same 
correlation-based methods used in \S \ref{sec:all_sky_analysis}, 
subtracting components correlated with Galactic tracers. 
For example, fitting the EGRB map as a sum of an isotropic flux $I_{iso}$ and 
a component linear in the $408\MHz$ tracer, gives $I_{iso}\simeq 0.86\pm0.19$. 
Using the synthetic large-scale structure tracer instead of the $408\MHz$ map 
yields $I_{iso}=0.55\pm0.35$. 

\begin{deluxetable}{cllc}
\tablecaption{\label{tab:Tracers} Galactic tracers}
\tablewidth{0pt}
\tablehead{\em Num. & \em Description & \em Prior & \em Ref.}
\startdata
1 & Low frequency synchrotron & $408\MHz$ all-sky survey & 1 \\
2 & High frequency synchrotron & $23\GHz$ WMAP foreground & 2 \\
3 & \HI column density & $21\cm$ all-sky map & 3 \\
4 & \HI column density & VTSS, SHASSA and WHAM $6563\angst$ maps & 4 \\ 
5 & Dust column density & COBE/DIRBE and IRAS/ISSA $100\mu$ maps & 5 \\
6 & Synthetic \HII column density & NE2001 model & 6 \\
7 & Large-scale Galactic structure & NE2001 model, 
global components only & 6 \\
\enddata
\tablerefs{
1. Haslam et al. 1982; 2. Bennett et al. 2003; 3. Dickey \& Lockman 1990; 
4. Finkbeiner 2003; 5. Schlegel et al. 1998; 6. Cordes \& Lazio 2002.}
\end{deluxetable}

In conclusion, the analysis of Sreekumar et al. was based on an elaborate 
model of Galactic \gama-ray emission, thus suffering from uncertainties 
regarding the distributions of the Galactic components, in particular CRs 
and radiation fields. 
Within present model uncertainties (see Appendix \ref{sec:galactic_model}), 
Galactic emission could account for the entire observed \gama-ray sky, 
precluding a robust identification of the EGRB. 
The EGRB anisotropy reported by Sreekumar et al. is inconsistent both with 
their assumption of isotropy and with an extragalactic origin. 
Moreover, their EGRB map is highly correlated with Galactic tracers, 
indicating that a large fraction of the reported EGRB flux is attributed 
to a Galactic foreground. 
It is difficult to estimate the level of extragalactic radiation present in 
this map, because the EGRB flux is low, the map has been significantly 
processed, and correlation-based method were shown in 
\S \ref{sec:all_sky_analysis} to provide only loose, upper limits on $I_x$.

\section{Discussion}
\label{sec:discussion}

We have analyzed the high energy \gama-ray data of EGRET in search for the 
illusive EGRB. 
The EGRB was identified by several previous studies, but with different, 
sometimes contradicting reported properties. 
We have investigated progressively larger parts of the \gama-ray sky, with 
increasingly stronger Galactic foreground concealing the EGRB. 
In contrast to previous studies, where the high-latitude data was found to 
include a large extragalactic contribution, we find no conclusive evidence for 
the existence of a diffuse isotropic \gama-ray background in any part of the 
sky. 
On the contrary, we find theoretical and phenomenological evidence for 
domination of the Galactic foreground at a level inconsistent with an 
isotropic EGRB as strong as previously reported. 
The \gama-ray data is found to agree with the expected signature of Galactic 
emission, with no need for an extragalactic component or a hypothetical 
emission halo surrounding the Galaxy, previously suggested as an alternative 
to an extragalactic origin \cite{Dixon98,Dar01a}. 

Since our analysis indicates that the extragalactic contribution to the 
\gama-ray sky is small, we are unable to infer its properties with the data 
presently available, but only derive an upper limit on its flux, $I_x$. 
The most reliable constraints are implied by the steep latitude-profile of 
the high latitude ($|b|>42\de$) regions, which exhibit strong Galactic features and agree well with the expected Galactic signature.  
These considerations yield $I_x<0.5\,(\at99\CL)$, but provide indications that 
the true flux is probably much smaller than this upper limit. 
We show that methods for identifying and subtracting the Galactic foreground 
by correlating the \gama-ray data with Galactic tracers, extensively used in 
previous studies, fail to properly eliminate this foreground, and are thus 
incapable of providing a reliable estimate of the EGRB properties. 
We reproduce the results of such studies by making the same assumptions, and 
demonstrate the sensitivity of the methods used to such assumptions. 
As an illustration, we show that by assuming the Galactic emission to follow 
the large scale structure of the Galaxy, such correlation-based methods give 
$I_x\la0.6$, lower than previous estimates by a factor $\ga2$. 

In our analysis we have focused on the spatial (angular) EGRET data, ignoring 
the spectral data, because of three reasons: 
(i) the EGRB is found to be much weaker than the Galactic foreground and modelling the spectral data can only lower our upper limit to the EGRB flux, 
(ii) a-priori information regarding the EGRB spectrum is poor, 
and (iii) there are poorly understood spectral variations in the Galactic 
foreground across the sky, even in regions of evident Galactic dominance 
\cite{Hunter97}. 
Furthermore, previous studies of the EGRET data have reported a hard EGRB spectrum with spectral indices $p\simeq 2.0-2.2$, similar to the expected spectrum of both the bremsstrahlung and the inverse-Compton components of the Galactic \gama-ray emission in energies around $100\MeV$ 
[see \S \ref{subsec:previous_correlation_studies}, Hunter et al. (1997), 
Sreekumar et al. (1998) and Appendix \ref{sec:galactic_model}]. 
For such an EGRB spectrum, the spectral information can not be used to 
distinguish between the EGRB and most of the Galactic foreground, further 
justifying our emphasis on the spatial dependence of the data. 

According to our study, virtually all the diffuse \gama-ray background 
measured by EGRET originates from within our Galaxy. 
A spectral break is observed between $100\MeV$ and $1\GeV$ in low latitudes, 
resulting from the decay of $\pi^0$ mesons produced in nucleon-nucleon 
scattering \cite[the ``pion bump'', e.g.][]{Hunter97}. 
This feature is stronger at low latitudes than at high latitudes, an effect 
previously partially attributed to a larger fractional contribution of the 
EGRB at high latitudes \cite{Hunter97, Sreekumar98}. 
The small extragalactic component implied by our analysis suggests that this 
effect directly reflects variations in the \gama-ray spectrum of the Galactic 
emissivity. 
Possible explanations include 
a large contribution from inverse-Compton scattering, as discussed in 
Appendix \ref{sec:galactic_model} \cite{Chi89,Dar01b}, 
a contribution from unresolved point sources such as pulsars 
\cite{Fatemi96,Weferling99}, 
and a decline in the CR proton-to-electron ratio with increasing distance from 
the Galactic plane \cite{Pohl91}. 
Interpretation of the \gama-ray data as directly reflecting Galactic emission, 
with insignificant extragalactic background even at high latitudes, 
should improve our understanding of the Galactic \gama-ray emission and of the 
components involved, in particular the spectrum and the distribution of CRs. 

The existence of a diffuse extragalactic \gama-ray background has been 
predicted by several studies, arising for example from unresolved blazars 
(e.g. Mukherjee \& Chiang 1999) 
or from CR electrons accelerated in intergalactic shocks, predominately around 
galaxy clusters \cite{Loeb00,Totani2000,Keshet03,Miniati02,Gabici03}. 
Since, according to our analysis, the EGRB is dominated by a much stronger 
Galactic foreground in every part of the sky, a future identification of the 
EGRB would require more elaborate techniques and better data than discussed 
here. 
With no temporal signature, and with a Galactic foreground of variable and 
poorly understood spatial and spectral behavior, \emph{direct} identification 
of the EGRB from the present data seems unlikely. 
The EGRB signal can be identified indirectly, by cross-correlating the 
\gama-ray data with relevant extragalactic tracers, such as galaxy counts, 
the X-ray background or radio emission from intergalactic shocks.
Scharf \& Mukherjee (2002) have cross-correlated the high latitude EGRET 
data with Abell clusters, identifying a positive signal at a confidence level 
of $3\sigma$, broadly consistent with the predictions of Keshet et al. (2003). 
Such methods could provide conclusive identification of an extragalactic 
diffuse \gama-ray signal with improved data, as expected from future \gama-ray 
missions such as the \emph{Gamma-Ray Large-Area Space Telescope} 
(GLAST\footnote{See http://glast.gsfc.nasa.gov}, planned to be launched in 
2006) and 
AGILE\footnote{See http://agile.mi.iasf.cnr.it/Homepage/index.shtml}, 
and \v{C}herenkov telescopes such as 
MAGIC\footnote{See http://hegra1.mppmu.mpg.de/MAGICWeb}, 
VERITAS\footnote{See http://veritas.sao.arisona.edu/} 
and HESS \footnote{See http://mpi-hd.mpg.de/hfm/HESS/HESS.html}.

The low flux of the EGRB inferred from our analysis imposes severe 
constraints on various astrophysical scenarios involving strong \gama-ray 
emission, such as 
\gama-ray emission from unresolved blazars and 
emission from intergalactic shocks. 
Summing up the expected contribution from unresolved blazars, 
$I_{blazars}=0.30_{-0.09}^{+0.10}$ \cite{Mukherjee99}, 
and the expected contribution from intergalactic shocks, 
$I_{shocks}\simeq 0.10_{-0.05}^{+0.10}$ \cite{Keshet03}, we obtain a diffuse 
extragalactic \gama-ray flux consistent, but not far from the upper limit 
found in our analysis. 
The proximity between the flux expected from unresolved blazars and 
intergalactic shocks and our upper limit on the EGRB flux, 
implies even more stringent limits on any additional extragalactic \gama-ray 
emitting processes. 

Finally, the new upper limit on the diffuse EGRB derived in this paper leads 
to revised, more stringent constraints on models for the production of high 
energy neutrinos and cosmic-rays. 
Waxman \& Bahcall (1999; Bahcall \& Waxman 2001) have shown that cosmic-ray 
observations set a model independent upper bound of 
$E_\nu^2\Phi_\nu<5\times10^{-8}\GeV\se^{-1}\cm^{-2}\sr^{-1}$ 
on the flux of high-energy neutrinos produced by $\gamma~p$ or $p~p(n)$ 
interactions in sources which are optically thin to high energy nucleons for 
these interactions (as is the case in both quasar jets and gamma-ray bursts), 
and which therefore contribute to the observed cosmic-ray flux. 
The answer to the question of whether or not the upper bound on the high 
energy neutrino flux can be evaded is important, since very large neutrino 
detectors are being designed for installation in the ocean or a deep lake 
\citep[see, e.g.][]{antares,nestor,baikal}, 
under Antarctic ice \citep[see e.g.][]{amanda}, 
in space \citep[see e.g.][]{owl1,owl2,owl3}, 
and using large area ground arrays \citep[see e.g.][]{auger1,auger2,watson}. 
There are two special types of sources for which the Waxman-Bahcall bound does 
not apply, and which could in principle produce a neutrino flux exceeding this 
limit. 
The first special type of source is one in which neutrinos are produced by 
processes other than photo-meson or proton-nucleon interactions; 
the second type is one for which the $\gamma~p$ or proton-nucleon optical 
depth is high. 
While we have no direct evidence, from either photon or high-energy cosmic-ray 
studies, to support the existence of such sources, constraints on the possible 
neutrino intensity produced by them is of interest for the design of high 
energy neutrino telescopes. The neutrino intensity from sources with high 
$\gamma~p$ or $p~p(n)$ optical depth for nucleons, but small optical depth for 
high energy photons, is constrained by the upper limit on the EGRB 
\citep[e.g.][]{Berezinskii90}. 
This is due to the fact that the production of charged pions, which decay to 
produce neutrinos, is accompanied by the production of neutral pions, which 
decay to produce high energy photons. 
The EGRB intensity of $I_x=1.45\pm0.05$ was used to infer an upper limit of 
$E_\nu^2\Phi_\nu<10^{-6}\GeV\se^{-1}\cm^{-2}\sr^{-1}$ on the neutrino 
intensity from such ``hidden'' sources 
(e.g. Mannheim, Protheore \& Rachen 2001). 
The reduced upper bound on the EGRB intensity obtained in this paper, 
$I_x\la 0.5$, implies a more stringent upper bound for ``hidden'' sources, 
$E_\nu^2\Phi_\nu<4\times10^{-7}\GeV\se^{-1}\cm^{-2}\sr^{-1}$, closer to the 
Waxman-Bahcall cosmic-ray bound. 
A smaller estimate of the EGRB flux, obtained by subtracting the calculated 
contributions of intergalactic shocks, would thus imply an even more stringent 
upper bound on such ``hidden'' sources. 

The detection of cosmic-rays of energy exceeding $10^{20}\eV$  
(see Nagano \& Watson 2000 for review) has led to extensive discussions of 
speculative models invoking new physics to account for the observed ultra-high 
energy cosmic-rays (see Bhattacharjee \& Sigl 2000 for review). 
While such exotic new physics is not required to account for the observed 
events with energies in excess of $10^{20}\eV$ \cite{Bahcall03}, it is 
worthwhile to note that the parameters of such models are constrained by the 
intensity of the EGRB (e.g. Sigl et al. 1997; Yoshida, Sigl \& Lee 1998; 
Bhattacharjee \& Sigl 2000 and references therein). 
The reduced upper limit on the EGRB derived here therefore implies more 
stringent constraints on new physics models for the production of ultra-high 
energy cosmic-rays.


\acknowledgments
\emph{Acknowledgments:}
We thank Robert Hartman, Stan Hunter, and Martin Pohl for valuable discussions. This work was supported in part by grants from NSF (AST-0204514 and AST-0071019) and NASA (NAG-13292), by a MINERVA grant and by a grant from the Rosa \& Emilio Segr\'e Fund (EW).


\appendix

\section{Appendix - Galactic \gama-ray Polar Foreground}
\label{sec:galactic_model}

In this Appendix we calculate the contribution of the Milky-Way galaxy to the 
\gama-ray intensity measured towards the Galactic poles, where the Galactic 
foreground is minimal. Such calculations rely on the current understanding of the Galactic components involved in high energy \gama-ray processes. Large uncertainties concerning these components render the calculated results inconclusive, permitting only order of magnitude estimates and imposing severe obstacles on any attempt to measure the EGRB by subtracting a modeled Galactic foreground. Our preceding analysis indicates an EGRB flux much lower than the Galactic foreground in any extended ($\ga 5\de$) patch in the sky, further complicating such attempts. 
This appendix does not attempt to present a sound model for Galactic \gama-ray emission, but rather reviews the present knowledge concerning the Galactic foreground and highlights its implications for the extragalactic background. Hence, although the Galactic model we use is similar to models used in other studies \cite{Bertsch93, Hunter97}, we emphasize the present uncertainties of the Galactic model and concentrate on estimates of the \emph{minimal} polar 
Galactic foreground implied. Such estimates have been used in \S \ref{sec:galactic_poles} to impose upper limits on the EGRB flux. 

Galactic high energy \gama-ray emission is believed to arise from interactions 
of CR electrons and protons with interstellar gas and radiation fields. 
The contribution of unresolved point sources appears to be small, according 
to the small contribution of resolved sources and the ``pion-bump'' feature 
apparent in the \gama-ray spectrum \cite{Hunter97}. 
In \S \ref{subsec:galactic_components} we present the vertical distributions 
(perpendicular to the Galactic plane) of the relevant Galactic components at 
the position of the sun in the Galactic plane. 
Next, we calculate the polar contribution of each \gama-ray emitting process: 
relativistic bremsstrahlung of CR electrons in the interstellar gas
(\S \ref{subsec:bremsstrahlung}), 
inverse-Compton scattering of various photon fields by CR electrons 
(\S \ref{subsec:inverse_compton}), 
and nucleon-nucleon scattering of CR protons with interstellar nuclei 
(\S \ref{subsec:nucleon_nucleon}). 
In \S \ref{subsec:total} we discuss the total \gama-ray polar brightness and 
general features of the high-latitude Galactic emission. 
The assumption made in this section are reviewed and examined in 
\S \ref{subsec:assumptions}. 
In \S \ref{subsec:final_conclusions} we summarize the main results of the 
calculation.

\subsection{Vertical Distribution of Galactic Components}
\label{subsec:galactic_components}

Here we discuss the distributions of interstellar gas, radiation fields and 
CR electrons and protons, at the position of the sun, perpendicular to the 
Galactic plane. 
The vertical distributions of these components, or at least their column 
densities towards the Galactic poles, are required for an estimate of the 
polar Galactic emission, and are summarized in Tables \ref{tab:polar} and 
\ref{tab:CR_distribution}.

\subsubsection{Gas}

At high latitudes, the column density of interstellar gas is dominated by 
\HImin. The column density of \HI towards the Galactic poles may be inferred from observations of various associated radiative processes. Using the \HI column density map based on $21\cm$ line emission (tracer 3), we obtain $N_{\tny{H I}}(|b|>86\de)=(1.4\pm0.1)\times 10^{20} \cm^{-2}$. After smoothing the map with the EGRET PSF-like Gaussian filter (see \S \ref{sec:data}), we find a higher column density, $N_{\tny{H I}}^{(s)} (|b|>86\de) = (1.6\pm0.2) \times 10^{20}\cm^{-2}$, because the observed latitude profile is steep (see \S \ref{subsec:gamma_ray_steepness}). The contribution of \HII to the polar column density, inferred from (either raw or smooth) \HII maps of a pulsar-based model (tracer 6), is $\sim 50\%$ smaller: $N_{\tny{H II}} (|b|>86\de) \simeq 0.8 \times 10^{20} \cm^{-2}$. No \HII complexes are observed towards the Galactic poles \cite{Paladini03}. The \HH contribution to the polar column density is smaller than the above two components, $N_{\mbox{\tiny{H}}_2} (|b|\ga 30\de) \la 0.2\times 10^{20} \cm^{-2}$ \cite{Dame01}, and we conservatively adopt for it a value of $\sim (0.1\pm0.1) \times 10^{20} \cm^{-2}$. Summing the above results yields a rough estimate of the polar column density of the gas. For example, the nucleon polar column density according to the raw (unsmooth) data is 
\beq \label{eq:proton_column_density} 
N_n(|b|>86\de)=(2.4\pm0.3)\times10^{20}\cm^{-2} \fin \eeq 
Although the polar column density of gas 
(eq. [\ref{eq:proton_column_density}]) has been roughly estimated, the gas 
distribution perpendicular to the Galactic plane is more uncertain. 
The vertical structure of gas at the Galactocentric position of the sun, 
dominated by the \HI structure, is often modeled as a combination of Gaussian 
and exponential functions, with an average FWHM of $230\pc$, but with strong 
fluctuations and internal structure on various scales \cite{Dickey90}.

\subsubsection{Radiation}

Three potentially important radiation fields are Compton-scattered by CR 
electrons to high energies, thus contributing to the Galactic \gama-ray 
emission: 
(i) the Galactic infrared background, (ii) the Galactic optical background, 
and (iii) the cosmic microwave background (CMB). 
The distributions of the IR and of the optical radiation fields through the 
Galaxy are uncertain, but may be estimated by integrating the Galactic 
emissivities obtained from models \cite{Bahcall80,Mathis83} or extracted, 
with some simplifying assumptions, from direct observations 
\cite{Boisse81,Kniffen81}. 
Kniffen \& Fichtel (1981) find, at the Galactocentric radius of the sun, an 
IR photon energy density that ranges from $U_{ir}(z=0)=0.61\eV\cm^{-3}$ to 
$U_{ir}(z=1\kpc)=0.45\eV\cm^{-3}$, 
where $z$ is the height above the Galactic plane, 
and similar optical energy densities: $U_{opt}(z=0)=0.67\eV\cm^{-3}$ 
and $U_{opt}(z=1\kpc)=0.34\eV\cm^{-3}$.  
Bloemen (1985) has used the model of Mathis et al. (1983) to obtain similar, 
although slightly lower, results. 
Chi \& Wolfendale (1991) find a similar optical energy density, ranging 
between $U_{opt}(z=0)=0.47\eV\cm^{-3}$ and $U_{opt}(z=1\kpc)=0.40\eV\cm^{-3}$, 
but a lower IR energy density: 
$U_{ir}(z=0)=0.27\eV\cm^{-3}$ and $U_{ir}(z=1\kpc)=0.20\eV\cm^{-3}$.  
Since the optical and infrared radiation fields have similar characteristic 
temperatures, $7\times10^3\K$ and $3.5\times10^3\K$ respectively, we use 
their combined energy density, roughly estimated to lie in the range 
$U_{tot}(z=0)=0.7-1.3\eV\cm^{-3}$ and $U_{tot}(z=1\kpc)=0.6-0.8\eV\cm^{-3}$. 
Hence, for any $|z|\la 1\kpc$, we may assume the radiation energy density to 
lie within the range 
\beq U_{tot}(|z|<1\kpc)=0.6-1.3 \eV\cm^{-3} \fin\eeq 
The uncertainty in the above estimates linearly affects the calculated 
inverse-Compton contribution to the Galactic polar \gama-ray emission, 
introducing a systematic error factor of order $2$ to the calculated 
inverse-Compton emissivity.

\subsubsection{Cosmic-ray Electrons}
\label{subsec:CR_electrons}

Since several processes, involving CR electrons of different energies, 
contribute to the polar \gama-ray foreground, our calculations depend on the 
CR distribution, both in energy and in height above the Galactic plane, $z$. 
Parameterizing the electron spectrum as a broken power-law, and assuming that 
the energy dependence is insensitive to $z$, we may write the vertical CR 
distribution at the position of the sun as 
\beq \label{eq:broken_power_law} 
\frac{dn}{d\gamma}(\gamma,z) = K(\gamma,z) \gamma^{-p(\gamma)} \coma \eeq
where $\gamma$ is the Lorentz factor of the electrons. 

\begin{figure}
\epsscale{1.0}
\plotone{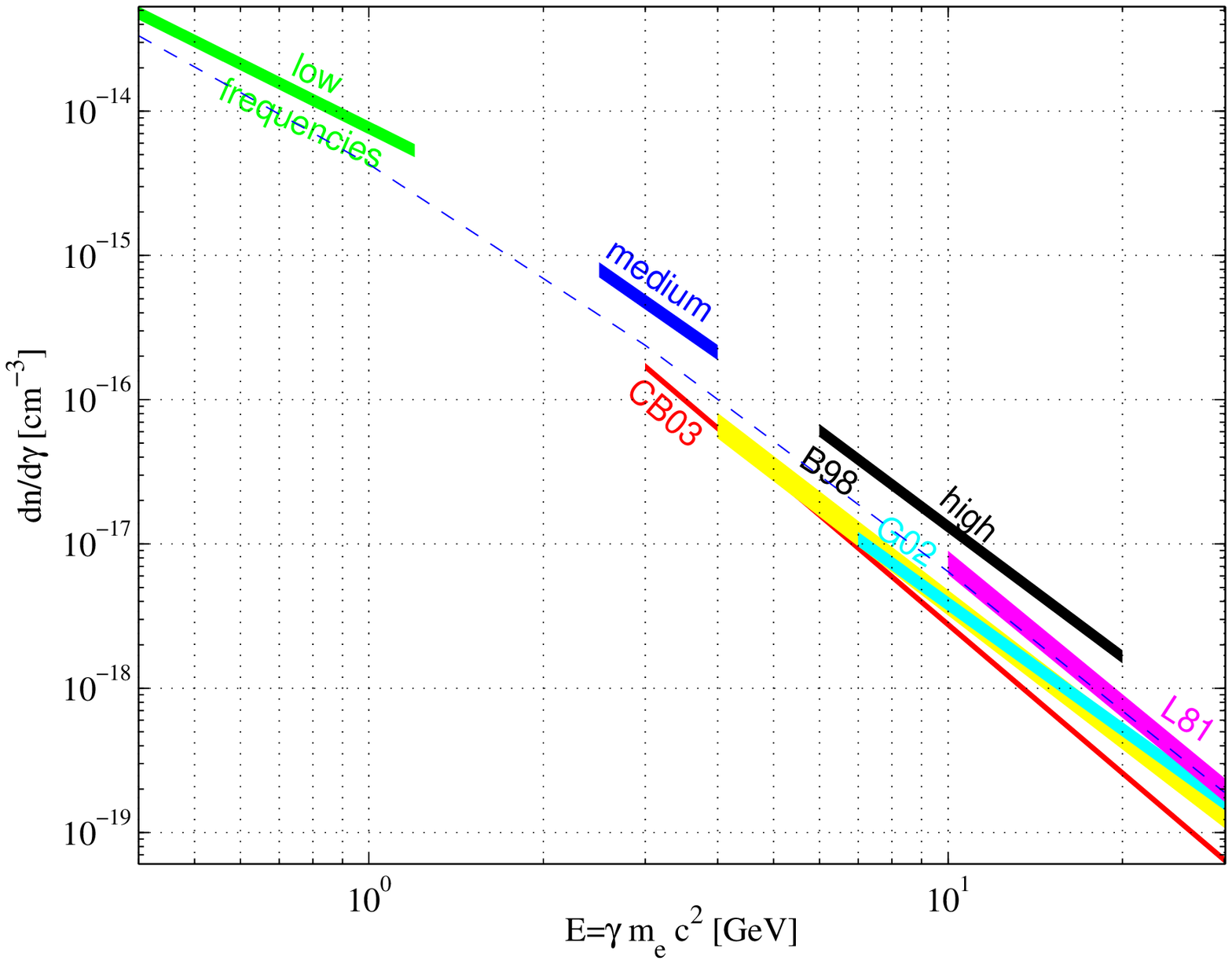}
\caption{ 
Reconstructed cosmic-ray electron distribution at the Galactocentric position 
of the sun (see equation [\ref{eq:broken_power_law}]). 
The results of direct measurements of the electrons arriving at the top of 
the atmosphere are presented, as adopted from 
Longair (1981, L81), 
Skibo \& Ramaty (1993, dashed line), 
Barwick et al. (1998, B98), 
Grimani et al. (2002, G02), 
and Casadei \& Bindi (2003, CB03). 
Also shown are the electron distributions inferred from radio observations 
towards the Galactic poles, at low ($\nu\la100\MHz$), 
medium ($400\MHz\la\nu\la1\GHz$) and high ($2\GHz\la\nu\la25\GHz$) 
frequencies, assuming an average magnetic field $B=5\muG$ and a scale height 
$L=1\kpc$. 
The thicknesses of the solid curves correspond to their reported $1\sigma$ 
confidence level or to their systematic error level. 
}
\label{fig:CR_distribution}
\end{figure}

At energies above $\sim10\GeV$, CR electrons are unaffected by the solar 
modulation, and their \emph{local} distribution is directly measured on top of 
the Earth's atmosphere. 
However, in spite of numerous observational efforts, significant discrepancies 
exist in the results and in their interpretation, leading to estimates that 
vary by a factor of $2-3$ around $10\GeV$, where counting statistics are 
considered good \cite{DuVernois01}. 
Figure \ref{fig:CR_distribution} shows typical estimates of the local CR 
electron distributions found by several studies. 
As a most recent example, Casadei \& Bindi (2003) have reported results 
well-fit for electrons of energies $3\GeV \la E \la 2\TeV$, by $p=3.42\pm0.02$ 
and $K(z=0)=(1.31\pm0.07)\times10^{-3}$. 
This flux is lower by a factor of $\sim 2.5$ (at $10\GeV$) than found by early 
studies \cite{Longair81}. 
Generally, some assumptions must be made when extrapolating the local electron 
distribution to other parts of the Galaxy. 
The vertical distribution of CRs in the Galactocentric position of the sun, 
needed for a calculation of the polar \gama-ray intensity, is often modeled as 
an exponent of scale height $\sim 1\kpc$, but such models are uncertain: the 
scale height may be different (e.g. Beuermann 1985) and vertical structure may 
be present. 

Information regarding the electron distribution in both $z$ and $\gamma$ may 
be indirectly deduced from radio observations of synchrotron radiation, 
emitted as the electrons gyrate in interstellar magnetic fields, 
with some assumptions regarding the strength and the distribution of magnetic 
fields through the Galaxy. 
In a randomly oriented magnetic field of amplitude $B$, the synchrotron 
emissivity of a power law electron distribution $dn/d\gamma=K\gamma^{-p}$ 
is given by \cite{Rybicki,Longair81}:
\begin{eqnarray} \label{eq:synch_spectrum} 
j_{syn}(\nu) & \simeq & 
\frac{\sqrt{3}}{2\pi} \frac{e^3 K B \sin\alpha}{m_e c^2} \left[
\Gamma(\frac{p}{4}+\frac{19}{12})\Gamma(\frac{p}{4}-\frac{1}{12})/(p+1)\right]
\left(\frac{2 \pi m_e c \nu}{3 e B \sin \alpha} \right)^{-\frac{p-1}{2}} \\
& \simeq & 1.04 \times 10^{-33} \,V(p) \left( \frac{K}{\cm^{-3}} \right) 
\left( \frac{B}{\muG} \right)^{\frac{p+1}{2}} 
\left( \frac{\nu}{\MHz} \right)^{-\frac{p-1}{2}}
\erg \se^{-1} \cm^{-3} \sr^{-1} \Hz^{-1} \coma \nonumber \end{eqnarray}
where an average over the pitch angle $\alpha$ was carried out in the second 
line, and the numerical coefficient $V(p)$, given by: 
\beq V(p) \equiv 1.35 \times 10^{-4} (2.9\times 10^{-3})^p 
\frac{ \Gamma(\frac{p}{4}+\frac{19}{12}) 
\Gamma(\frac{p}{4}-\frac{1}{12}) \Gamma(\frac{p+5}{4})}
{(p+1) \Gamma(\frac{p+7}{4})} \coma \eeq
is well approximated by $V(1.5\la p\la 3.5) \simeq 700^{(2.5-p)}$ 
in the relevant range of spectral indices. 
For electron energies where the CR spectrum (eq. [\ref{eq:broken_power_law}]) 
is well approximated by a single power law, 
i.e. where $K(\gamma,z)$ and $p(\gamma)$ depend only weakly on $\gamma$, 
equation (\ref{eq:synch_spectrum}) may be used to estimate the synchrotron 
emissivity at the corresponding frequency, 
\beq \nu_{syn}(E) \label{eq:synch_frequency} \simeq 
\frac{3}{16} \frac{e B}{m_e c} \left(\frac{E}{m_e c^2}\right)^2 
\simeq 1 \left(\frac{B}{\mu \mbox{G}}\right) 
\left( \frac{E}{\mbox{282 MeV}}\right)^2 \MHz \fin \eeq 
Conversely, for frequencies $\nu$ where the measured radio spectrum is a 
power-law, one may use equation (\ref{eq:synch_spectrum}) to reconstruct the 
electron distribution at the corresponding energy $E$. 
Thus, a photon spectral index $s$ implies $p(E)=2s+1$, and the measured 
emissivity $j_\nu$ gives a crude estimate of the flux normalization: 
\beq \label{eq:elec_dist} 
K(E)=9.5\times 10^{32} \, V(2s+1)^{-1}
\left(\frac{j_\nu}{\mbox{erg}\se^{-1}\cm^{-3}\sr^{-1}}\right) 
\left(\frac{B}{1\mu\mbox{G}}\right)^{-(s+1)} 
\left(\frac{\nu}{1\MHz}\right)^{s} \cm^{-3} \fin \eeq

For the relevant spectral indices, $s=0.5-1.0$, the CR flux normalization 
$K(E)$ depends strongly on the magnetic field amplitude $B$. 
The interstellar magnetic field has been extensively studied from synchrotron 
emission, Faraday rotation, polarization of optical starlight and Zeeman 
splitting. 
Typical estimates of the total magnetic field in the solar neighborhood, 
including a regular component aligned along the Galactic spiral structure, 
and an irregular, randomly oriented component, range between $4\muG$ and 
$6\muG$ \cite{Beck96,Widrow02}.  
In the following we use a consensus value $B=5\muG$ \cite{Han01}, but retain 
the dependence of the results on magnetic field. 

Radio observations of the Galactic polar regions in frequencies 
$6\MHz \la \nu \la 100\MHz$ reveal a power-law spectrum, of spectral index 
$s\simeq0.5$ \cite{Cane79}. 
This implies that the electron spectrum may be approximated as a power law of 
spectral index $p\simeq2.0$ in the corresponding range of electron energy, 
$300\MeV \la E \la 1.2\GeV$. 
The synchrotron emissivity at $10\MHz$ has been estimated based on radio 
observations towards opaque \HII regions \cite{Fleishmann95} as 
$j(10\MHz)\simeq(3.0\pm0.3)\times 10^{-39}\junits$. 
The same emissivity is obtained using the effective radio temperature of the 
Galactic poles, $T(\nu=10\MHz)\simeq (3.0\pm0.2)\times 10^5\K$ \cite{Cane79}, 
assuming a uniform synchrotron emissivity of scale height $L\simeq 1\kpc$ 
\cite{Longair81}. 
We may now find the average normalization of the electron distribution in the 
relevant electron energies from equation (\ref{eq:elec_dist}): 
$\langle K\rangle=(2.9\pm0.3)\times 10^{-8} (B/5\muG)^{-1.5}\cm^{-3}$.
At higher frequencies, $400\MHz \la \nu \la 1\GHz$, the radio spectral index 
of the poles changes to $s\simeq 0.9$ \cite{Sironi74,Webster74}, implying a 
steepening of the electron spectrum to $p=2.8$ at the corresponding energies, 
$2.5\GeV \la E \la 4\GeV$. 
The synchrotron emissivity at such frequencies may be obtained, as for the 
lower frequencies, from the measured polar brightness, by assuming 
$L\simeq 1\kpc$. 
The emissivity thus estimated, e.g. from the $408\MHz$ synchrotron map 
(tracer 1), gives, according to equation (\ref{eq:elec_dist}), 
$\langle K\rangle\simeq1.7\times10^{-5}(L/1\kpc)^{-1}(B/5\muG)^{-1.9}\cm^{-3}$.
Comparing the $408\MHz$ and the $23\GHz$ polar data gives a spectral index 
$s\simeq1.0$ \cite{Bennett03}, indicating that the spectrum slightly steepens 
at some frequency above $1\GHz$, e.g. in the range  $2\GHz\la\nu\la25\GHz$. 
This implies that the electron spectrum in energies $6\GeV\la E\la 20\GeV$ 
has a spectral index $p\simeq 3.0$, with normalization 
$\langle K\rangle\simeq9.9\times10^{-5}(L/1\kpc)^{-1}(B/5\muG)^{-2.0}\cm^{-3}$ 
found from the $23\GHz$ polar brightness (tracer 2).   
Systematic errors of $\la10\%$ may be found in the radio data used for the 
latter two estimates. 
The CR electron distribution averaged towards the poles, as inferred above 
from synchrotron emission, is summarized in table \ref{tab:CR_distribution}.

\begin{deluxetable}{lcc}
\tablecaption{\label{tab:CR_distribution} Average Cosmic-ray Electron Distribution Based on Synchrotron Emission}
\tablewidth{0pt}
\tablehead{\em Energy Range & \em $K \,[\mbox{cm}^{-3}]$ \tablenotemark{a} & $p$ \tablenotemark{a} }
\startdata
$300\MeV\la E\la 1.2\GeV$ & $(2.9\pm0.3)\times 10^{-8}(B/5\muG)^{-1.5}$ & $2.0$ \\
$2.5\GeV\la E\la 4\GeV$ & $1.7\times10^{-5}(L/1\kpc)^{-1}(B/5\muG)^{-1.9}$ & $2.8$ \\
$6\GeV\la E\la20\GeV$ & $9.9\times10^{-5}(L/1\kpc)^{-1}(B/5\muG)^{-2.0}$ & $3.0$ \\
\enddata
\tablenotetext{a}{The average density of electrons with Lorentz factor 
$\gamma$, averaged towards the Galactic poles, is parameterized as 
$dn(\gamma)/d\gamma \simeq K \gamma^{-p}$.}
\end{deluxetable}

Figure \ref{fig:CR_distribution} shows the reconstructed CR electron 
distribution at the Galactocentric position of the sun, featuring both direct, 
local measurements carried out on top of the Earth atmosphere, 
and indirect estimates of an average CR flux, based on the polar radio 
brightness and assuming $B=5\muG$ and $L=1\kpc$. 
The figure illustrates that the reconstructed spectrum gradually slopes 
down at higher energies, with rough agreement between the estimated flux at 
different energies. 
However, local direct measurements yield an electron flux \emph{lower} than 
inferred from radio observations, by a factor of $\sim 3$ at $E=4\GeV$, and by 
a factor that ranges between $\sim1.5$ and $\sim5$ at $E=10\GeV$, because 
different experiments disagree on the flux of CR electrons arriving on top of 
the Earth atmosphere. 
There are two possible explanations for the difference between the results of 
the two methods \cite{Longair81}. 
First, the CR flux near the sun may not be representative of the average flux 
at the Galactocentric position of the sun, e.g. if the CR flux is dominated by 
single sources \cite{Erlykin98} or if the sun is located in a low density 
region, where the CR electron flux is lower than in adjacent regions situated 
towards the Galactic poles. 
Observations suggest that the sun is approximately in the center of a low 
density region of gas (the local hot bubble), extending $\sim 0.3\kpc$ towards 
each pole \cite{Cordes02}, and correlations expected between gas and CRs 
(on an unknown scale, e.g. Hunter et al. 1997) may suggest the same for CRs. 
Alternatively, the electron flux inferred from radio observations may have 
been overestimated by a factor of $\sim3$, if the Galactic magnetic field or 
the scale height of synchrotron emission were underestimated. 
In particular, note that the reconstructed electron flux is sensitive to the 
magnetic field amplitude $B$ and to its fluctuations along the line of sight, 
because the measured radio intensity is proportional to the integral of the 
electron distribution weighted by $B^{(s+1)}$. 

The preceding discussion indicates that in spite of numerous efforts, the CR 
electron distribution towards the Galactic poles remains highly uncertain. 
Only two observables, measured and interpreted at limited accuracy, are 
available: the local CR flux arriving on top of the Earth atmosphere, and a 
combination of CR flux and magnetic field amplitude integrated along the line 
of sight. 
With little information regarding the vertical structure of the CRs and the 
magnetic fields, the simplest plausible assumptions are made (an exponential 
distribution of scale height $L=1\kpc$ and $B=5\muG$ on average),
but the implied differences between the two observables suggest that the 
true situation is different and perhaps more complicated. 
Hence, the results obtained in this section should be considered as order of 
magnitude estimates only, accurate at best to within a factor of $\sim2-3$. 
Consequently, similarly large systematic errors are also found in all present 
calculations of the Galactic \gama-ray emission.  
The implications of these uncertainties are discussed in 
\S \ref{subsec:assumptions}. 

\subsubsection{Cosmic-ray Protons}
\label{subsec:CR_protons}

The local flux and spectrum of CR protons have been measured with high 
precision on top of the Earth atmosphere, at energies above $1\GeV$ per 
nucleon, where modulation by the solar wind has little effect 
\cite{Boezio99,Wang02}.
The distribution of CR protons through the Galaxy, however, is not directly 
probed, and requires some simplifying assumptions. 
One possibility is to assume that the CR proton-to-electron ratio is uniform 
through the Galaxy, thus inferring the CR proton distribution from the 
distribution of CR electrons, as deduced for example from synchrotron 
emission.  
Alternatively, some authors have assumed that the CR proton distribution is 
correlated with the distribution of gas, and modeled it by convolving the gas 
distribution with a Gaussian filter function of HWHM $r_0$ corresponding to 
some coupling scale \cite{Hunter97}. 
For our purpose, only the vertical distribution of the CR protons at the 
Galactocentric position of the sun, and only along the distribution of the 
gas, is of importance. 
It should be noted, however, that a uniform CR proton-to-electron ratio 
suggests that the uncertainty regarding the CR electron flux averaged towards 
the poles, as discussed in \ref{subsec:CR_electrons}, may also apply to the CR 
proton distribution. 
Namely, in analogy with the CR electron distribution, the CR proton flux 
averaged at the Galactocentric position of the sun may be higher by a factor 
of order $3$ than locally measured.

\subsection{Bremsstrahlung}
\label{subsec:bremsstrahlung}

An important contribution to the Galactic \gama-ray emission arises from 
relativistic bremsstrahlung of CR electrons with the interstellar plasma. 
A relativistic electron of energy $E=\gamma m_e c^2$, propagating in a plasma 
of characteristic ISM composition ($\sim0.9$ hydrogen and $\sim0.1$ helium by 
number), thus emits radiation with specific photon number emissivity 
approximated by \cite{Longair81} 
\beq \frac{dj}{d\epsilon}(\epsilon,E) \simeq \frac{1}{4\pi} 
\frac{\alpha\,n_{gas}}{\epsilon} \Theta(E-\epsilon) \coma\eeq
where $\alpha \simeq 10^{-15}\cm^3 \se^{-1}$ and the step function $\Theta$ 
ensures that photons of energy $\epsilon$ are emitted only by electrons of 
higher energy $E$. 
The difference between screened and unscreened bremsstrahlung is small, of the 
order of $20\%$, and irrelevant for the order of magnitude estimate used here. 
In order to derive the emissivity of an electron distribution 
$dn/d\gamma=K \gamma^{-p}$ above a threshold photon energy 
$\epsilon_0=100\MeV$, one must integrate over the electron distribution and 
over the photon energy: 
\begin{eqnarray} \label{eq:brem} \nonumber j_{brem}(>\epsilon_0) & = & 
\int_{\epsilon_0}d\epsilon \,\frac{dj(\epsilon)}{d\epsilon}(\epsilon,E) 
\int_{\epsilon/m_e c^2}d\gamma \, \frac{dn}{d\gamma}(\gamma) \\
& \simeq & \frac{1}{4\pi}\frac{\alpha\,n_{gas}\,K}{(p-1)^2} 
\left( \frac{\epsilon_0}{m_e c^2} \right)^{-(p-1)} \fin \end{eqnarray}
Most of this radiation is emitted by electrons with energies close to (but 
higher than) $100\MeV$. 
For example, if $p=2$, electrons of energies $100\MeV\leq E\la1\GeV$ are 
responsible for $90\%$ of the emission above $100\MeV$. 

At energies $300\MeV \la E \la 1.2\GeV$, the electron distribution was 
reconstructed in \S \ref{subsec:CR_electrons} from radio observations, 
giving $p\simeq 2.0$ and 
$\langle K\rangle \simeq (2.9\pm0.3) \times 10^{-8}(B/5\muG)^{-1.5} \cm^{-3}$. 
At lower energies, $100\MeV \leq E \la 300\MeV$, the electron distribution is 
inaccessible because the corresponding synchrotron frequencies $\la 6\MHz$ are 
too low to be observed on earth, but the electron spectrum is likely similar. 
In order to find the Galactic bremsstrahlung contribution to the polar 
intensity, we integrate equation (\ref{eq:brem}) over the line of sight: 
\beq \label{eq:int_brem} I(>100\MeV) \simeq (1.2\pm 0.1)\times 10^{-26} 
\left[ \int_0 \frac{K(z)}{\langle K\rangle} \frac{n_{gas}(z) \,dz}{ 
\mbox{cm}^{-2}} \right]\ph \cm^{-2} \se^{-1} \sr^{-1} \fin \eeq 
As mentioned in \ref{subsec:CR_electrons}, the vertical structure of the CR 
electron distribution at the Galactocentric position of the sun is uncertain. 
If we \emph{assume} that the CR flux is nearly constant across the gas 
distribution (with FWHM $\sim230\pc$), then the integral in 
equation (\ref{eq:int_brem}) approximately equals the column density of gas, 
$N_n$, and the resulting polar \gama-ray intensity (above $100\MeV$, 
in $10^{-5} \Iunits$ units) is 
\beq I_{brem} = (0.28\pm0.04) \left(\frac{B}{5\muG}\right)^{-1.5}\fin \eeq
The actual intensity depends also on the nature of correlations between 
CRs and gas: a positive correlation may enhance the above result considerably, 
whereas an anti-correlation could, in principle, eliminate it.
In light of the direct measurements discussed in \S \ref{subsec:CR_electrons}, 
the CR flux used above may have been overestimated by a factor of $\sim 3$, 
as is the case if the scale height of CR and magnetic fields is $\sim 3\kpc$, 
yielding $I_{brem}\simeq 0.1$.

\subsection{Inverse-Compton}
\label{subsec:inverse_compton}

An electron of Lorentz factor $\gamma\gg 1$ may inverse-Compton scatter a 
photon of low energy $\epsilon_i$ to a high average energy 
$\epsilon_f \simeq (4/3)\gamma^2 \epsilon_i$, assuming that the scattering in 
the electron rest frame is approximately elastic \cite{Rybicki}.
When the low energy photons are thermal with a characteristic temperature $T$, 
the average energy of a scattered photon is 
$\epsilon_f \simeq 3.6 \gamma^2 k_B T$, implying that CR electrons of 
characteristic energy $E\simeq 300\, (T/1\K)^{-1/2} \GeV$ are required in 
order to scatter such a thermal photon up to $\sim 100\MeV$ energies. 
With the electron distribution in the relevant energy range approximated as 
a power law, $dn/d\gamma\simeq K \gamma^{-p}$, the specific emissivity of 
inverse-Compton scattering a blackbody photon field \cite{Rybicki} may be 
integrated to give:
\begin{eqnarray} \label{eq:IC_spectrum} 
j_{iC}(>\epsilon_0) & \simeq & 
\frac{2 Q(p)}{p-1} c \sigma_T K U_{ph} (k_B T)^{\frac{p-3}{2}} 
\epsilon^{-\frac{p-1}{2}} \\ 
& \simeq &  2.1 \times 10^{-15} \frac{\widetilde{Q}(p)}{(p-1)}
\left(\frac{K U_{ph}/k_B T}{\mbox{cm}^{-6}} \right)
\left(\frac{\epsilon}{k_B T}\right)^{-\frac{p-1}{2}} 
\ph \se^{-1} \cm^{-3} \sr^{-1} \coma \nonumber \end{eqnarray} 
where $Q(p)$ is a numerical coefficient, defined by: 
\beq Q(p) \equiv \frac{45}{4\pi^5} \frac{2^p (p^2+4p+11)}{(p+3)^2(p+5)(p+1)}  
\Gamma\left(\frac{p+5}{2}\right) \zeta \left(\frac{p+5}{2}\right) \coma\eeq
$\zeta(x)$ is the Riemann zeta function, 
and $\widetilde{Q}(p) \equiv Q(p)/Q(3) = 6\pi Q(p)$ is well approximated in 
the relevant range of spectral indices by 
$\widetilde{Q}(2.5\la p\la 3.5) \simeq 1+0.93(p-3)+0.51(p-3)^2$. 

Inverse-Compton scattering of CMB photons up to $\sim 100\MeV$ energies 
requires $\sim 180\GeV$ electrons.  
Estimates of the electron distribution at these energies, based on direct 
measurements carried out on top of the atmosphere, were presented above, 
although shown to vary between different studies. 
For a CMB temperature $T_{cmb}=2.73\K$ and energy density 
$U_{cmb}=0.26 \eV \cm^{-3}$, we find from equation (\ref{eq:IC_spectrum}) 
inverse-Compton emissivities in the range 
$j_{iC}(>100\MeV)\simeq(1.6-6.1)\times 10^{-29}\ph\se^{-1} \cm^{-3} \sr^{-1}$. 
Assuming that the CR flux measured locally is representative of a smooth 
vertical distribution of scale height $1\kpc$, we find that inverse-Compton 
scattering of CMB photons contributes little to the polar \gama-ray intensity: 
\beq I_{iC,cmb}\simeq 0.005-0.019 \fin \eeq 
This result is insensitive to details of the vertical CR distribution, 
because the CMB distribution is smooth. 
The contribution of CMB scattering to the Galactic \gama-ray intensity remains 
small, $I_{iC,cmb}\la 0.06$, even if the CR flux used above was underestimated 
by a factor of $3$, e.g. because the sun lies in a low CR density region. 

Inverse-Compton scattering of optical and IR photons, of characteristic temperatures $7\times 10^3\K$ and $3.5\times 10^3 \K$, respectively, is dominated by $\sim 4\GeV$ CR electrons. 
The distribution of such electrons was estimated from radio observations at 
the corresponding frequencies, $\sim 1\GHz$, as $p\simeq 2.8$ and 
$\langle K\rangle=(1.7\pm0.2)\times 10^{-5}(L/1\kpc)^{-1}(B/5\muG)^{-1.9} 
\cm^{-3}$.  
If we \emph{assume} that the vertical distributions of photons and CR 
electrons are smooth, then the scale height $L$ cancels out when we integrate 
the emissivity towards the Galactic pole, giving a polar \gama-ray intensity: 
\beq
I_{iC}(>100\MeV) \simeq 0.35 \times 10^{-5} \frac{\langle U_{ph}\rangle}
{\mbox{eV}\cm^{-3}} \left(\frac{T}{5\times 10^3\K}\right)^{-0.1} 
\left(\frac{B}{5\muG}\right)^{-1.9} \ph \se^{-1} \cm^{-2} \sr^{-1} \coma \eeq
where $\langle U_{ph}\rangle$ is the energy density of radiation averaged 
along the CR distribution. 
This result depends very weakly on the temperature of the photon field, 
justifying treatment of the optical and the IR fields on the same footing, 
but is linear in the energy density of the fields. 
Estimates of the optical and the IR energy densities discussed above, in the 
range $0.6-1.3 \eV\cm^{-3}$, thus lead to 
\beq I_{iC} = (0.2-0.5) \left(\frac{B}{5\muG}\right)^{-1.9} \eeq
This result should be treated as an order of magnitude estimate only; 
in addition to the assumptions mentioned above, the true intensity is 
proportional to the integrated product of CR flux and photon field energy 
density, and thus sensitive to the distribution of the two and to the 
correlations among them. 
Hence, strong correlations between Galactic photons and CRs will enhance the 
resulting polar intensity significantly, whereas anti-correlation between the 
two may, in principle, eliminate the inverse-Compton component. 
Another potential systematic error lurks, of course, in the CR flux used 
above: 
if it was overestimated by a factor of $\sim 3$, a possibility discussed in 
\S \ref{subsec:CR_electrons}, then the inverse-Compton contribution could 
reduce to $I_{iC}=0.07-0.15$.

\subsection{Nucleon-Nucleon Scattering}
\label{subsec:nucleon_nucleon}

Collisions between CR protons and the interstellar gas produce pions, 
which in turn decay and emit \gama-rays, mostly through neutral $\pi^0$ decay. 
The resulting emissivity depends on the density of gas and on the distribution 
of CR protons. 
The emissivity per interstellar hydrogen atom has been calculated by several 
authors, using the local spectrum and flux of CR protons, measured on top of 
the Earth atmosphere. 
Dermer (1986) reviews several such calculations, finding most results in the 
range 
\beq n_{gas}^{-1}\,j_{\pi}(>100\MeV)=(1.0-1.4)\times10^{-26}
\ph \se^{-1} \sr^{-1}\mbox{ atom}^{-1} \fin \eeq 
Assuming that the CR proton distribution is approximately constant along the 
distribution of gas, with the flux and the spectrum measured \emph{locally}, 
integration of the emissivity towards the Galactic pole becomes trivial. 
For the median estimated emissivity per atom, 
$1.2\times10^{-26}\ph \se^{-1} \sr^{-1}\mbox{ atom}^{-1}$, 
we find that nucleon-nucleon scattering contributes 
$I_{\pi}\simeq 0.28\pm0.04$ to the Galactic polar \gama-ray intensity. 
A minimal estimate of Galactic nucleon-nucleon intensity towards the pole may 
be obtained using minimal estimates of both the gas column density, 
$N_n\ga 2.0\times 10^{20}\cm^{-2}$, and of the emissivity per atom. 
Such a minimal estimate, combined with a similarly obtained maximal estimate, 
yield
\beq 0.20 \la I_{\pi}\la 0.36 \eeq

The nucleon-nucleon scattering component of the polar \gama-ray intensity 
along the line of sight could be significantly larger than estimated above. 
First, correlations between CR protons and the gas could enhance the emission 
considerably. 
Second, since the emissivity was calculated based on the CR proton flux 
measured locally, a higher average flux towards the pole would also produce a 
higher \gama-ray intensity. 
Such a possibility is advocated by an analogy with the CR electron 
distribution, if the CR proton-to-electron ratio is uniform at the 
Galactocentric position of the sun (see \S \ref{subsec:CR_protons}). 
Consequently, the emissivity could be higher by a factor of $\sim 3$, implying 
$0.6\la I_{\pi}\la 1.1$.

\subsection{Total Galactic \gama-ray Polar Foreground}
\label{subsec:total}

Thus far we have obtained order of magnitude estimates of the three processes 
dominating the Galactic high-energy ($>100\MeV$) \gama-ray emission at the 
Galactocentric position of the sun: 
bremsstrahlung by CR electrons in the interstellar gas, inverse-Compton 
scattering of the optical and IR radiation fields by CR electrons, 
and nucleon-nucleon scattering of CR protons in the interstellar gas. 
Estimates of these three contributions to the polar \gama-ray intensity and of 
the associated Galactic components, are summarized in table \ref{tab:polar}. 
We find that the three processes have comparable contributions to the polar 
\gama-ray intensity, our calculations being limited by present uncertainties 
concerning the distributions of the relevant Galactic components,
most importantly the Galactic CR electrons and protons. 
Summing up the \emph{median} estimates of these three processes yields a 
median estimate of the Galactic polar intensity: $I_{gal} \simeq 0.9$. 
Taking into account the uncertainties inherent to the calculation, in terms 
of $1\sigma$ probability intervals where such are available, gives 
\beq  I_{gal}=0.6-1.2 \fin \eeq 
However, systematic errors in the underlying assumptions permit a broader 
possible range of polar intensities. 
The dependence of $I_{gal}$ on the assumptions made is discussed in 
\S \ref{subsec:assumptions}. 

As mentioned in \S \ref{subsec:high_latitude_fit}, the Galactic \gama-ray 
emission is expected to be correlated with tracers of various Galactic 
components. 
Thus, the inverse-Compton component of the Galactic \gama-ray emission should 
be correlated with synchrotron maps that reflect the distribution of CR 
electrons, but it is impossible to predict the extent of such correlations 
with available information regarding the three-dimensional distributions of 
radiation and of magnetic fields through the Galaxy. 
The bremsstrahlung and the nucleon-nucleon components of the Galactic 
\gama-ray emission are expected to be correlated with tracers of the 
interstellar gas, as well as with radio maps, but the strengths of these 
correlations, which depend also on the three-dimensional cross-correlations 
between gas and cosmic-rays, are unknown. 
Hence, although the total Galactic \gama-ray emission should be correlated 
with such tracers, it is generally impossible to predict the strength of these 
correlations and many other features of the \gama-ray sky, 
because of the limited information regarding the three-dimensional 
distributions of the Galactic components and their cross-correlations. 

Under some conditions, global features of the \gama-ray sky, such as its 
average latitude dependence, would be approximately proportional to a linear 
combination of a synchrotron map (with weight $w_{syn}$) and a tracer of the 
gas column density (with weight $w_{gas}=1-w_{syn}$). 
This would be the case, for example, if fluctuations on the relevant scales 
in the CRs or in the gas column density are weak, and the magnetic field 
distribution is smooth. 
The weights of such a linear combination, $w_{syn}$ and $w_{gas}$, are 
a-priori unknown and could vary across the sky. 
However, at high latitudes, where the composition of Galactic components 
varies slowly and local features are rare, these weights should be 
approximately constant. 
The two  weights can be crudely estimated, under extensive assumptions 
regarding the distribution of Galactic components. 
For example, if we assume that the distributions of radiation and of magnetic 
fields through the Galaxy are smooth, then the inverse-Compton component of 
the Galactic emission is expected to be proportional to the synchrotron maps. 
If, in addition, we assume that fluctuations in both CRs and gas are small on 
the relevant scales (as is the case, for example, when studying the average 
latitude profile of the Galactic tracers at very high latitudes), then the 
bremsstrahlung and the nucleon-nucleon components should be proportional to 
the \emph{normalized sum} of a radio map and a gas tracer. 
The ratio $w_{syn}:w_{gas}$ could vary between $6:4$ and $8:2$ under these 
assumptions.

\begin{deluxetable}{lcc}
\tablecaption{\label{tab:polar} Galactic Polar Estimates}
\tablewidth{0pt}
\tablehead{\em Component & \em Value \tablenotemark{a} & \em Units }
\startdata
\HI column density & $1.4\pm 0.1$ & $10^{20} \cm^{-2}$ \\
\HII column density & $0.8$ & $10^{20} \cm^{-2}$ \\
\HH column density & $0.1\pm 0.1$ & $10^{20} \cm^{-2}$ \\
Nucleon column density & $2.4\pm 0.3$ & $10^{20} \cm^{-2}$ \\
\tableline
IR energy density \tablenotemark{b} & $0.2-0.6$ & $\mbox{eV}\cm^{-3}$ \\ 
Optical energy density \tablenotemark{b} & $0.3-0.7$ & $\mbox{eV}\cm^{-3}$ \\ 
Total (IR + optical) \tablenotemark{b} & $0.6-1.3$ & $\mbox{eV}\cm^{-3}$ \\ 
\tableline
Effective $408\MHz$ temperature & $14\pm1$ & K \\
Effective $23\GHz$ temperature & $87\pm9$ & $\mu\mbox{K}$ \\
\tableline
Bremsstrahlung scattering \tablenotemark{c} & $(0.29\pm 0.04)(B/5\muG)^{-1.5}$ & $10^{-5}\Iunits$ \\
Inverse Compton scattering \tablenotemark{c} & $(0.2-0.5)(B/5\muG)^{-1.9}$ & $10^{-5}\Iunits$ \\
Nucleon-nucleon scattering \tablenotemark{c} & $0.20-0.36$ & $10^{-5}\Iunits$ \\
\tableline
Measured \gama-ray intensity & $1.20 \pm0.08$ & $10^{-5}\Iunits$ \\
Calculated \gama-ray intensity \tablenotemark{c} & $0.6-1.2$ & $10^{-5}\Iunits$ \\
\enddata
\tablenotetext{a}{When based on a tracer: for $|b|>86\de$ before smoothing.}
\tablenotetext{b}{For height $0<z<1\kpc$ above the Galactic plane.}
\tablenotetext{c}{Calculated based on the assumptions summarized in 
\ref{subsec:assumptions}.}
\end{deluxetable}

\subsection{Assumptions and their Validity}
\label{subsec:assumptions}

The above range of estimates for the polar \gama-ray intensity, 
$I_{gal}=0.6-1.2$, reflects only the relatively mild uncertainties, 
associated mostly with the distributions of gas and radiation fields. 
Potentially much larger systematic errors lurk in the assumptions made, 
typical of such models for Galactic \gama-ray emission: 
\begin{enumerate}
\item{The CR electron distribution is based on radio observations of 
synchrotron emission. }
\item{The CR proton distribution is inferred from measurements on top of the 
atmosphere. }
\item{Magnetic fields of amplitude $5\muG$ are assumed at the solar 
Galactocentric position. }
\item{The vertical CR and magnetic field distributions are smooth, 
and have a scale height of $L=1\kpc$. }
\item{No correlations between CRs and gas or radiation are assumed towards the 
poles. }
\end{enumerate}

Possibly large systematic errors lie in the assumed flux of CR electrons and 
protons. 
In \S \ref{subsec:CR_electrons} we presented evidence that the local CR 
electron flux, measured on top of the atmosphere, is lower than radio-based 
estimates of the average flux towards the pole, by a factor of $\sim 3$, 
and possibly in the range $\sim 1.5-5$, for the assumed values of $B$ and $L$. 
This could result from misinterpretation of the radio data, implying an 
average CR electron flux lower than assumed, or conversely, indicate that the 
locally measured flux is not representative of the average 
(e.g. if the CR flux is dominated by single sources or if the sun is located 
in a low CR density region, see \S \ref{subsec:CR_electrons}), 
possibly implying an average flux of CR protons higher than assumed 
(see \S \ref{subsec:CR_protons}). 
The bremsstrahlung and inverse-Compton components of the \gama-ray emission 
are both linear in an average flux of CR electrons 
(but averaged differently for each component), whereas the nucleon-nucleon 
scattering component is linear in an averaged flux of CR protons. 
Thus, if the CR electron flux is lower by a factor of $\sim 3$ than used in 
the calculation, but the CR proton flux is close to its locally measured 
value, we find $0.4\la I_{pole}\la 0.6$, where $> 50\%$ of the emission arises 
from nucleon-nucleon scattering.  
A CR proton-to-electron flux ratio higher than this would imply Galactic 
emission with spectral \gama-ray features associated with pion decay 
(the ``pion bump''), stronger than observed at low latitudes \cite{Hunter97}, 
where Galactic domination is overwhelming. 
Hence, the CR electron flux is not likely to be lower than used in our 
calculation by a factor larger than $\sim3$, implying a minimal polar 
intensity $I_{pole}\ga0.4$.  
Alternatively, a CR electron flux corresponding to radio observations and an 
average CR proton flux higher than measured locally, would imply a polar 
\gama-ray intensity higher than observed, dominated by nucleon-nucleon 
scattering. 

The calculated bremsstrahlung and inverse-Compton components are sensitive to 
the magnetic field amplitude, through its effect on the interpretation of the 
synchrotron data; 
stronger magnetic fields imply weaker \gama-ray emission.  
Magnetic fields of amplitude $B\sim 10\muG$ could resolve the disagreement 
between the CR electron flux measured locally and the flux inferred from 
radio observations. 
An amplitude much larger than $6\muG$ would disagree with various measurements 
of the interstellar magnetic field, although the parameter relevant for 
inferring the CR electron flux from radio observations, 
$B^{1+s}$ (where $s=0.5-1.0$), is sensitive to fluctuations along the line of 
sight and is not directly measured. 
Considering as an extreme possibility a magnetic field of amplitude $B=8\muG$ 
(when averaged towards the pole according to $B^{1+s}$), 
while leaving the other assumptions unchanged, yields $0.4\la I_{pole}\la0.7$. 

The calculated bremsstrahlung and inverse-Compton \gama-ray components are 
similarly sensitive to the scale height of CR electron distribution, $L$. 
The average CR electron flux deduced from radio observations, 
$\langle K\rangle$, scales as $L^{-1}$, implying similar scaling of the 
bremsstrahlung component because the scale height of gas distribution is much 
smaller than $L$. 
The inverse-Compton component dependence upon $L$ is likely weaker, because 
the unknown scale height of radiation is probably larger than the scale height 
of gas, possibly of order $L$.   
CR scale heights as large as $2\kpc$ at the Galactocentric position of the 
sun have been inferred from synchrotron emission models \cite{Beuermann85}, 
but values larger than $3\kpc$ are unlikely, because the implied radio-based 
CR electron flux would then be lower than measured locally. 
For $L=3\kpc$, the two methods of estimating the CR electron flux roughly 
agree, with consequent polar \gama-ray intensity $0.4\la I_{pole} \la 0.7$. 

Since the \gama-ray emission of the Galaxy arises from interactions between 
the different Galactic components, the resulting \gama-ray intensity depends 
on correlations, of various scales, between these components. 
Hence, positive correlations between CRs and gas or radiation fields may 
enhance the resulting \gama-ray intensity significantly, 
whereas anti-correlations could eliminate most of the calculated intensity. 
Little is known about the correlations of CRs with the Galactic components, 
thus imposing large uncertainties on any calculation of Galactic \gama-ray 
emission. 
On average, however, one does not expect to find strong 
\emph{anti}-correlations between these Galactic components, suggesting that 
lower limits on the \gama-ray polar intensity obtained above are robust.

\subsection{Concluding Remarks}
\label{subsec:final_conclusions}

Two major conclusions of this section are worth highlighting. 
First, in spite of great progress in the understanding of processes leading 
to Galactic \gama-ray emission and of the Galactic components involved, 
calculations of the Galactic \gama-ray emission remain considerably uncertain. 
Thus, although the measured Galactic \gama-ray emission can be modeled 
\cite{Bertsch93,Hunter97}, such models have not yet reached the accuracy 
required in order to subtract the strong Galactic foreground and measure the 
weaker extragalactic background \cite{Sreekumar98}. 
Second, although the model uncertainties are large, a robust lower limit may 
nevertheless be imposed on the Galactic contribution to the \gama-ray 
intensity measured towards the poles. 
We find that even extreme, yet plausible assumptions, yield calculated 
Galactic contributions to the polar intensity \emph{no lower} than 
$I_{min}\simeq 0.4$. 
More conventional assumptions give higher polar intensities, estimated in the 
range $I_{gal}= 0.6-1.2$. 
Our results are in accord with previous calculations of the polar Galactic 
foreground \cite{Fichtel78}.



\end{document}